\documentclass[]{rsos}%%%%where rsos is the template name

\usepackage[super,sort&compress,comma]{natbib}

\usepackage{wasysym}
\usepackage{marvosym}

\begin{document}

\title{Dust evolution, a global view: III. Core/mantle grains, organic nano-globules, comets and surface chemistry}

\author{%%%% Author details
A. P. Jones}  

%%%%%%%%% Insert author address here
\address{Institut d'Astrophysique Spatiale, CNRS, Univ. Paris-Sud, Universit\'e Paris-Saclay, B\^at. 121, 91405 Orsay cedex, France}

%%%% Subject entries to be placed here %%%%
\subject{Astrochemistry, Astrophysics}

%%%% Keyword entries to be placed here %%%%
\keywords{Interstellar medium, interstellar dust, interstellar molecules}

%%%% Insert corresponding author and its email address}
\corres{A.P. Jones\\
\email{Anthony.Jones@ias.u-psud.fr}}

%%%% Abstract text to be placed here %%%%%%%%%%%%
\begin{abstract}
Within the framework of {\em The Heterogeneous dust Evolution Model at the IaS} (THEMIS) this work investigates in detail the surface processes and chemistry relating to core/mantle interstellar and cometary grain structures and its influence on the nature of these fascinating particles. 
It appears that a realistic consideration of the nature and chemical reactivity of interstellar grain surfaces could self-consistently and within a coherent framework explain:  
the anomalous oxygen depletion, 
the nature of the CO dark gas, 
the formation of "polar ice" mantles, 
the red wing on the 3\,$\mu$m water ice band, 
the basis for the O-rich chemistry observed in hot cores,  
the origin of organic nano-globules and
the $\sim 3.2\,\mu$m "carbonyl" absorption band observed in comet reflectance spectra. 
It is proposed that the reaction of gas phase species with carbonaceous a-C(:H) grain surfaces in the interstellar medium, in particular the incorporation of atomic oxygen into grain surfaces in epoxide functional groups, is the key to explaining these observations. 
Thus, the chemistry of cosmic dust is much more intimately related with that of the interstellar gas than has previously been considered. The current models for interstellar gas and dust chemistry will therefore most likely need to be fundamentally modified to include these new grain surface processes.  
\end{abstract}
%%%%%%%%%%%%%%%%%%%%%%%%%%%

\maketitle

%------------------------------------------------------------------
\section{Introduction}
\label{sect_intro}
%------------------------------------------------------------------

Interstellar dust has been something of a problem for more than 80 years, ever since the early measurements of interstellar reddening by Trumpler.\cite{1930PASP...42..214T} The first dust models to attempt to explain this extinction followed about a decade or so later and perhaps the earliest viable proposition was the dirty ice model of van de Hulst.\cite{vandehulst43} Soon after, the idea that these dirty ice particles would evolve in the interstellar medium (ISM) was considered and their life-time was estimated to be about 50 million years.\cite{1946BAN....10..187O} Some thirty years later dust modelling became more sophisticated with the consideration of graphite, enstatite, olivine, silicon carbide, iron and magnetite particles as viable dust materials that could be used to explain interstellar extinction.\cite{1977ApJ...217..425M} This study concluded that graphite was a necessary dust component for any viable dust model and that it could be combined with any of the other materials to satisfactorily match the observed extinction. From these early studies were born the now widely-accepted notion that interstellar dust principally consists of graphite and some form of silicate. So, for the last few decades graphite and amorphous silicate materials have therefore formed the basis of the most widely used dust models, which have indeed served us well.\cite{1984ApJ...285...89D,2001ApJ...551..807D,2001ApJ...554..778L,2007ApJ...657..810D}  However, recent observational evidence shows that this approach is no longer a satisfactory or sufficient basis for a realistic dust model.\cite{2015A&A...577A.110Y,2015A&A...580A.136F}
Some recent work has re-visited our long-held views on the nature of dust in the ISM and is, hopefully, forcing us to re-examine some of our widely-held views about it.\cite{2012A&A...540A...1J,2012A&A...540A...2J,2012A&A...542A..98J,2013A&A...555A..39J,2013A&A...558A..62J,JonesTaiwanDustPaper,2014A&A...565L...9K,2014P&SS..100...26J,Faraday_Disc_paper_2014,2015A&A...579A..15K,2015A&A...000A.000J}
Unsurprisingly dust has therefore become a much more complex, and therefore an inherently more interesting, subject of study of late. 

This work is an attempt to provide a global framework within which to link and interpret, in a self-consistent and coherent manner, a wide diversity of open questions within the sphere of interstellar dust research. These relate to the observational and laboratory analyses of:
\begin{itemize}
\item carbon depletions and their variations, 
\item C-shine (cloud- and core-shine), 
\item the oxygen depletion problem, 
\item the CO dark gas mystery, 
\item interstellar and proto-stellar ices, 
\item organic nano-globules and  
\item comet reflectance spectra. 
\end{itemize}
The following therefore principally revolves around the links between dust, chemistry and dust surface chemistry. 
ISM studies of the role and importance of the gas-dust interaction have long-held to the classical view of dust as a passive surface on which to combine hydrogen atoms to form molecular hydrogen and to hydrogenate heavier species which are then ejected into the gas to drive and enrich the gas phase chemistry. 
However, as emphasised here, dust surfaces are far from passive and probably play a key role in chemistry in space. 
For, without dust in the ISM there would be very little chemistry going on there and without the molecular coolants resulting from that chemistry there would be little star formation, few planets and little likelihood of life. Also, without chemistry there would indeed be no dust.

The finer details of the ideas presented here will almost certainly be shown to be wrong in the course of time. 
However, it is hoped that the global approach to cosmic dust evolution that is proposed here will provide a framework within which future studies can be developed to test these ideas. 

The paper is organised as follows: 
Section \ref{sect_precepts} discusses carbon depletion and accretion from the gas, the nature of core/mantle grains, C-shine and mantle evolution, 
Section \ref{sect_DustI} considers the consequences of dust evolution, accretion anomalies, surface chemistry and surface epoxide and aziridine functional groups, grain surface carbonyl groups and CO sequestration from the gas, 
Section \ref{sect_DustII} discusses the role of evolved grain mantles, "organic" materials and nano-globules, "volatile ice" mantles, photolysis effects, hot core processing, comets and chemistry, 
Section \ref{sect_tests} suggests some experiments that might be used to explore the ideas proposed here and
Section \ref{sect_conclusions} concludes this work.

%------------------------------------------------------------------
\section{Dust: some basic precepts}
\label{sect_precepts}
%------------------------------------------------------------------  

At the heart of the ideas and the global approach presented here is the new  dust modelling framework THEMIS (The Heterogeneous dust Evolution Model at the IaS).\cite{2013A&A...558A..62J,2014A&A...565L...9K,2015A&A...000A.000J,2015A&A...000A.000Y,2016A&A...999A..99J} This framework explicitly assumes that, through the effects of dust evolution, interstellar grains must be rather well mixed and comprise chemically-distinct materials incorporated or transformed into core/mantle (CM) grains and more complex aggregate structures in dense clouds.  Perhaps the most important and overriding aspect of this modelling is that it assumes that the dust composition and structure evolve in response to the local physical conditions (gas temperature and density, radiation field, turbulence, shocks, \ldots) through the various effects of erosion, irradiation, accretion, coagulation {\it etc}.  

The following sub-sections provide a series of dust chemistry anchor points for the following sections that have been  developed within, and now extend, the THEMIS framework.

%------------------------------------------------------------------
\subsection{Core/mantle (CM) grains and beyond (CMM, AMM, AMMI)}
\label{sect_precepts_CMs}
%------------------------------------------------------------------  

The long-standing idea of core/mantle or CM interstellar grains\cite{1986Ap&SS.128...17G,1997A&A...323..566L} was recently given a new treatment.\cite{2013A&A...558A..62J,2014A&A...565L...9K} This recent development, the THEMIS dust modelling framework,\cite{2016A&A...999A..99J} is built upon a new core/mantle model for dust in the diffuse ISM and the evolution of the dust properties in response to their local environment.\cite{2013A&A...558A..62J,2014A&A...565L...9K,Faraday_Disc_paper_2014,2014A&A...570A..32B,2015A&A...577A.110Y,2015A&A...000A.000J,2015A&A...000A.000Y} The underlying principle of the THEMIS modelling is the supposition that interstellar dust is not the same everywhere but that it evolves within a given region of the interstellar medium (ISM) as it reacts to and interacts with its local environment. For example, photon, ion and electron irradiation can induce changes in the dust chemical composition and structure, and accretion and coagulation will drive changes in the grain structure. All of these processes directly affect the dust optical properties, which are the key to understanding the nature of dust. 
In the tenuous ISM the outer carbonaceous layers of the grains, be they carbon grains or the mantles on other grains, will be H-poor and aromatic rich due to UV photolysis by stellar FUV/EUV photons.\cite{2012A&A...540A...1J,2012A&A...540A...2J,2012A&A...542A..98J,2013A&A...558A..62J,ANT_RSOS_nanoparticles} The innermost parts of carbon grains, shielded by an optically-thick outer a-C layer or mantle, can retain any original a-C:H core or could possibly attain it through H atom interaction and hydrogenation of the grains inner regions. Thus, the large grains in the ISM are likely to have a-C:H/a-C and a-Sil/a-C core/mantle (CM) structures. 
The THEMIS model therefore supposes carbon-coated amorphous silicate grains with iron and iron sulphide nano-particle inclusions (a-Sil$_{\rm Fe,FeS}$/a-C), and core/mantle amorphous carbon grains (a-C:H/a-C). The model is built upon the laboratory-measured properties of interstellar dust analogue materials, {\it i.e.}, amorphous silicates, iron, iron sulphide and hydrogenated amorphous carbons materials, and provides a viable explanation for the observed interstellar dust IR-FUV extinction, IR-mm thermal emission, dust absorption/emission spectra and the evolution of the dust properties in the transition between diffuse  and molecular regions.\cite{2013A&A...558A..62J,2014A&A...565L...9K,Faraday_Disc_paper_2014,2015A&A...000A.000J,2015A&A...000A.000Y}

The THEMIS modelling of dust evolution in the transition from tenuous to denser regions\cite{2012A&A...548A..61K,Faraday_Disc_paper_2014,2015A&A...579A..15K,2015A&A...000A.000J,2015A&A...000A.000Y} predicts that all grains in the outer reaches of molecular clouds, where the UV radiation field is significantly attenuated, will be coated with a second mantle to form core/mantle/mantle grains (CMM) through the accretion of carbon from the gas phase as a-C:H (H-rich and aliphatic-rich hydrocarbons) and the coagulation of the a-C nano-particles onto grain surfaces (see the following sub-sections \ref{sect_precepts_Cshine} and \ref{sect_mantle_evolution}). Deeper into a cloud these grains will coagulate into aggregates (AMM), which can then accrete ice mantles (AMMI). 

As has already been pointed out\cite{Faraday_Disc_paper_2014} the large a-C:H/a-C grains of the THEMIS model 
qualitatively resemble the interesting and intriguing organic nano-globules extracted from meteorites, interplanetary dust particles (IDPs) and cometary dust (see Section \ref{sect_DustII}). Indeed, if organic nano-globules were in part of interstellar origin then it is to be expected that there ought also to be a population of  the nano-globules with silicate cores. The analysis of a supernova silicate grain did reveal the presence of a $\sim 25$\,nm surface layer of organic material,\cite{2015LPI....46.1609D} which would appear to lend some support to the core/mantle interstellar grain hypothesis. 
However, given that in the THEMIS model, the carbon mantles on the silicate grains are so much thinner ($\simeq 5-10$\,nm) than the photo-processed surfaces layers ($\simeq 20-30$\,nm) of the carbonaceous (a-C:H/a-C) grains, the search for a-C mantles on silicates will be intrinsically more difficult.

%------------------------------------------------------------------
\subsection{Carbon depletion and accretion}
\label{sect_precepts_depletion_accretion}
%------------------------------------------------------------------ 

Carbon is an extremely important element in the ISM because it is actively implicated in both gas and solid phase chemistry, in the former case carbon, in its singly-ionised form C$^+$ or C\,II, is an important interstellar gas coolant. In the latter case in the form of hydrogenated amorphous carbon, a-C(:H), dust particles. The large family of amorphous carbons encompasses materials from those rich in hydrogen and aliphatic (sp$^3$) carbon, a-C:H, to those poor in hydrogen and rich in aromatic (sp$^2$) carbon, a-C, which all come under the collective label of a-C(:H). 

A good determination of the total abundance of carbon in the ISM, the stock of all carbonaceous gas and dust species, has always been problematic because the C\,II transitions useful for absorption-line studies are in the UV and are either very strong or very weak,\cite{2012ApJ...760...36P} thus hampering the determination of a key parameter for ISM gas and dust studies. 
Despite early indications that there was a carbon crisis, {\it i.e.}, that most dust models required more carbon than was apparently cosmically available,\cite{1995Sci...270.1455S} it now appears that interstellar carbon is sufficiently abundant to fulfil all the functions that are required of it.\cite{2012ApJ...760...36P}

%------------------------------------------------------------------
\subsubsection{Carbon depletion}
\label{sect_precepts_depletion}
%------------------------------------------------------------------ 

Besides being rather more abundant than previously thought it also appears that the gas and dust phase abundances of carbon are rather variable, indicating that dust undergoes significant processing in the atomic ISM and that there is a significant exchange of carbon between the gas and dust phases.\cite{2012ApJ...760...36P} Hence, the amount of carbon locked up in dust, the carbon depletion, would seem to reflect the local conditions, particularly the local gas density. 
Certain observable characteristics of interstellar dust can be directly attributed to a-C(:H) nano-particles, {\it i.e.}, the FUV extinction rise, the UV bump centred at 217\,nm and the $3-13\,\mu$m IR thermal emission bands from stochastically-heated particles. 
It is therefore to be expected that variations in the carbon depletion ought to be reflected in associated changes in these observables or vice versa.
However, it appears that the intensity of the UV extinction bump does not correlate with the depletion of carbon into dust.\cite{2012ApJ...760...36P}  
This apparent paradox can be explained by cloud geometry effects because depletion into dust is driven by accretion in dense cloud interiors, {\it i.e.}, by cloud volume, but the UV bump intensity and the FUV extinction reflect the abundance of small grains, {\it i.e.}, a-C(:H) nano-particles, in the lower density surface regions of clouds,\cite{2013A&A...558A..62J} {\it i.e.}, a "skin" effect as is observed in the case of the diffuse interstellar bands (DIBs). 
It has also been noted that the carbon depletion and the FUV extinction show a gradual decrease with decreasing gas density, which could be due to the UV-driven photo-fragmentation of small grains and the liberation of carbon into the gas in the diffuse ISM.\cite{2012ApJ...760...36P,2012A&A...542A..69P,2013A&A...558A..62J} 
Further, the HD 207198 line of sight shows a very high carbon depletion into dust and a weak UV bump,\cite{2012ApJ...760...36P} which in association a strong FUV extinction is characteristic of the a-C:H materials expected to accrete in the outer regions of dense clouds.\cite{2013A&A...558A..62J,2015A&A...000A.000J,2015A&A...000A.000Y}

Given the large fraction of cosmic carbon that is likely locked up in a-C(:H) nano-particles,\cite{2013A&A...558A..62J} and the apparent susceptibility of a-C(:H) nano-particles to UV photo-processing in the tenuous ISM,\cite{2012A&A...542A..69P,2012A&A...542A..98J,2013A&A...558A..62J} it is perhaps not surprising that interstellar carbon depletions show wide variations that are seemingly coupled with systematic variations in the UV extinction characteristics along a given line of sight.

%------------------------------------------------------------------
\subsubsection{Carbon accretion}
\label{sect_precepts_accretion}
%------------------------------------------------------------------ 

% FIGURE 1
\begin{figure}[!h]
\centering\includegraphics[width=3.5in,angle=270.]{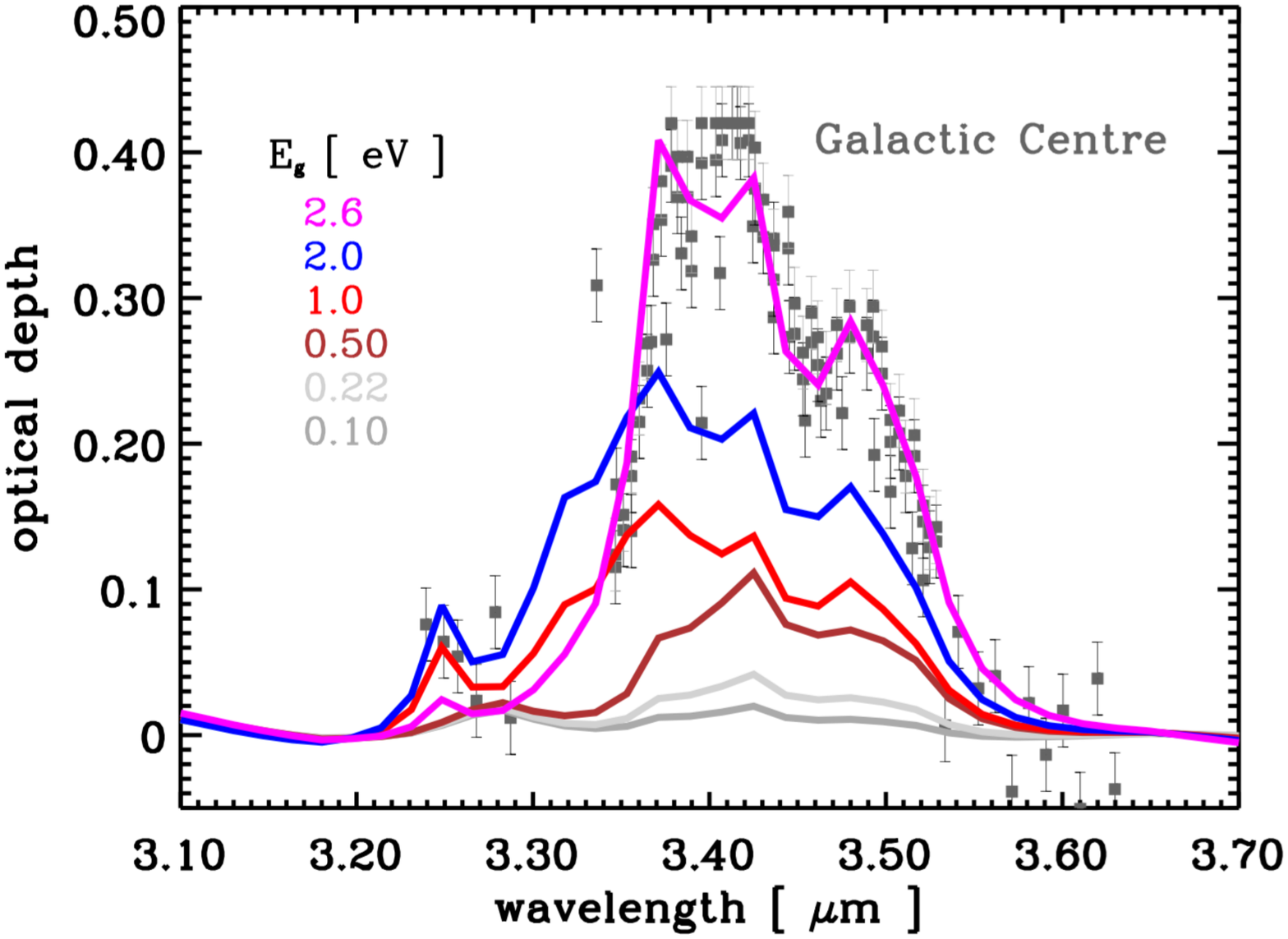}
\centering\includegraphics[width=3.5in,angle=270.]{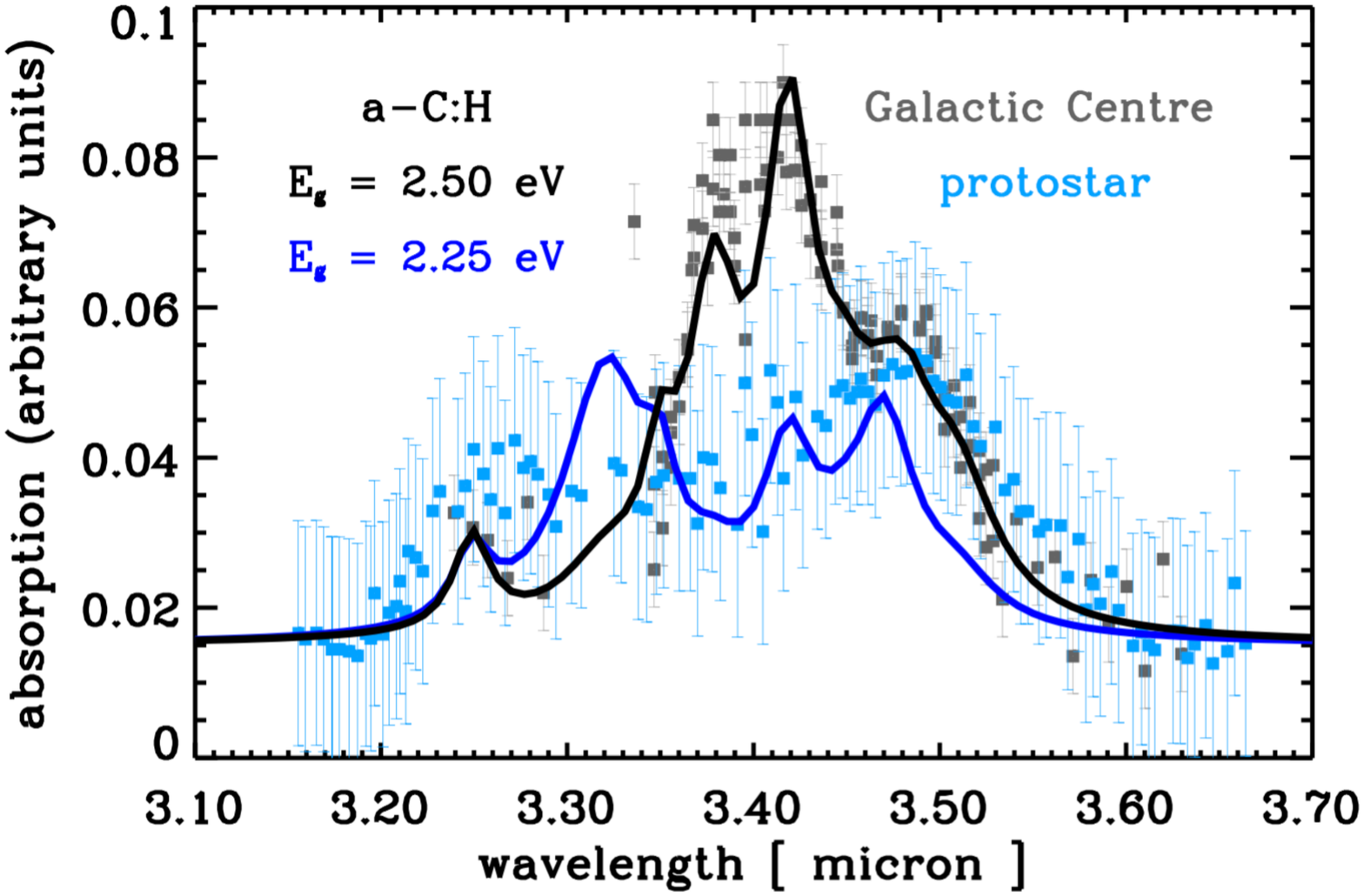}
\caption{Upper plot: The $3-4\,\mu$m region continuum-subtracted optical depth data as a function of the core/mantle (CM) grain a-C(:H) mantle band gap for $E_{\rm g} = 2.6$\,eV (violet), $2.0 $\,eV (blue), $1.0 $\,eV (red), $0.5 $\,eV (brown), $0.22 $\,eV (light grey) and $0.1 $\,eV (grey). Also shown are the scaled-to-fit spectra of the Galactic Centre towards IRS6E and Cyg OB2 No. 12 (grey squares).\cite{2002ApJS..138...75P} 
Lower plot: Spectra of the Galactic Centre towards IRS6E and Cyg OB2 No. 12 (grey squares)\cite{2002ApJS..138...75P} and that re-determined towards the protostar Mon~R2/IRS-3 (blue squares).\cite{1994ApJ...433..179S} The black and blue lines show the spectrum for aliphatic-rich a-C:H materials with $E_{\rm g} = 2.5$ and $2.25 $\,eV, respectively.}
\label{fig_3mic_ext}
\end{figure}

The accretion or formation of a-C(:H) mantles in the transition from tenuous to denser interstellar media has important and observable consequences for the dust observed in these transition regions.\cite{2012A&A...548A..61K,2015A&A...579A..15K,2015A&A...000A.000J,2015A&A...000A.000Y}
The absorption spectra of a-C(:H) materials in the $3.2-3.6$\,$\mu$m wavelength region show broad features composed of about ten sub-features due to aliphatic, olefinic and aromatic C-H stretching bands, of which more than half are due to aliphatic CH$_n$ ($n = 1,2,3$) bonds. 
The exact form of this absorption band is therefore a function of the a-C(:H) composition (H atom content and band gap, $E_{\rm g}$)\footnote{As per previous work\cite{2012A&A...540A...2J} we adopt the Tauc definition of the band gap, $E_{\rm g}$, for a-C(:H) materials.} and structure (C atom sp$^3$/sp$^2$ bonding ratio) and in the ISM its form will vary with the degree of photo-processing and/or hydrogenation experienced there.\citep{2012A&A...540A...1J,2012A&A...540A...2J,2012A&A...542A..98J,2013A&A...558A..62J} 
The upper plot in Fig.~\ref{fig_3mic_ext} shows the continuum-subtracted, dust model spectrum in the $3.1-3.7$\,$\mu$m region for the THEMIS CM  grains\cite{2013A&A...558A..62J} as a function of the a-C(:H) mantle band gap, $E_{\rm g} = 0.1-2.6$\,eV. 
For comparison this figure also shows the absorption spectrum for the diffuse ISM towards the Galactic Centre (grey squares), which indicates the presence of wide band gap a-C:H dust materials along this line of sight. 
The broad nature of the $3.2-3.7$\,$\mu$m feature and its complex structure makes for problematic baseline-subtraction  because of the often limited wavelength coverage of the observations and also because of preconceived notions about the shape of the continuum and the nature of the bands.\cite{JonesTaiwanDustPaper}  

\begin{figure}[!h]
\centering\includegraphics[width=3.5in,angle=270.]{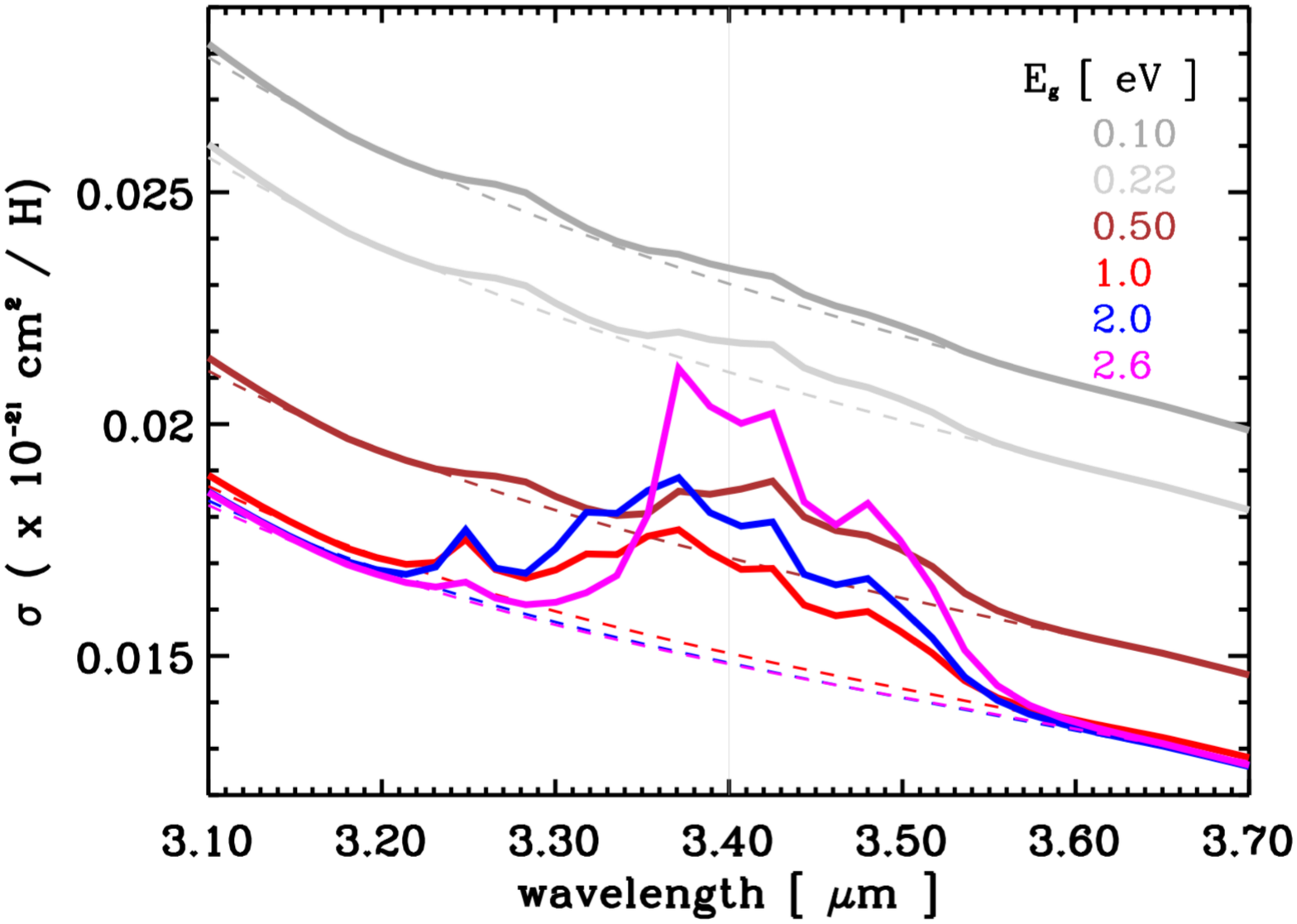}
\caption{The dust model extinction cross-section as a function of the outer a-C(:H) mantle band gap.\cite{2013A&A...558A..62J}  The dashed lines show the adopted 4$^{\rm th}$ order polynomial baselines fitted to the model data at $\lambda =$ 3.20 and 3.65\,$\mu$m.}
\label{fig_3mic_crosssect}
\end{figure}

Interstellar absorption spectroscopy in the $3.2-3.6\,\mu$m wavelength region of dense clouds towards protostars and through the Taurus molecular cloud \citep{1992ApJ...399..134A,1994ApJ...433..179S,1995ApJ...449L..69S} revealed what appeared to be some interesting and unusual features at  3.25 and 3.47\,$\mu$m. Of these two features the 3.47\,$\mu$m feature was attributed to a tertiary C-H stretch on diamond.\cite{1992ApJ...399..134A}  
However, the analysis of the spectra in this region\cite{1994ApJ...433..179S,1995ApJ...449L..69S} relied upon the subtraction of rather {\it ad hoc} baselines that may have skewed the interpretation of these data by removing a significant "plateau" from underneath the bands.\cite{JonesTaiwanDustPaper} 
For example, in the case of Mon~R2/IRS-3, the adoption of a non-ideal but more realistic linear baseline, rather than an assumed polynomial baseline,\cite{1994ApJ...433..179S,1995ApJ...449L..69S} results in a broad absorption band with relatively weak superimposed features, which looks more like a typical amorphous hydrocarbon spectrum than that deduced by the authors \citep{1994ApJ...433..179S,1995ApJ...449L..69S,JonesTaiwanDustPaper}. Hence, it would appear that the long-held notion that the 3.4\,$\mu$m feature, typical of the diffuse ISM, disappears in molecular clouds may be erroneous.\citep{JonesTaiwanDustPaper} It is rather that the shape of the C-H stretching band towards denser molecular regions is different, and so the key question here is rather, how different is this band?
The lower plot in Fig.~\ref{fig_3mic_ext}  shows the model\cite{2013A&A...558A..62J} spectra (solid lines) compared to that for the diffuse ISM towards the Galactic Centre (grey squares)\cite{2002ApJS..138...75P} and the re-baselined spectrum\cite{JonesTaiwanDustPaper} towards the protostar Mon~R2/IRS-3 (blue squares).\cite{1994ApJ...433..179S}  
In the latter spectrum a linear baseline was subtracted from the observational data rather than the polynomial used by the authors.\cite{1994ApJ...433..179S} This broad plateau absorption is qualitatively consistent with a-C:H materials with band gap $E_{\rm g} \simeq 2$\,eV and is typical of the material is expected to form by accretion in the denser, molecular regions of the ISM where little UV processing of the mantle material is possible.\citep{Faraday_Disc_paper_2014,2015A&A...000A.000J,2015A&A...000A.000Y}
In  Fig.~\ref{fig_3mic_crosssect} the full dust model IR absorption spectra are shown,\cite{2013A&A...558A..62J} these are the continuum-included spectra from Fig.~\ref{fig_3mic_ext} for a-C(:H) mantle materials with $E_{\rm g} = 2.6$\,eV (violet), $2.0 $\,eV (blue), $1.0 $\,eV (red), $0.5 $\,eV (brown), $0.22 $\,eV (light grey) and $0.1 $\,eV (grey). 

% FIGURE 2

\begin{figure}[!h]
\centering\includegraphics[width=3.5in,angle=270.]{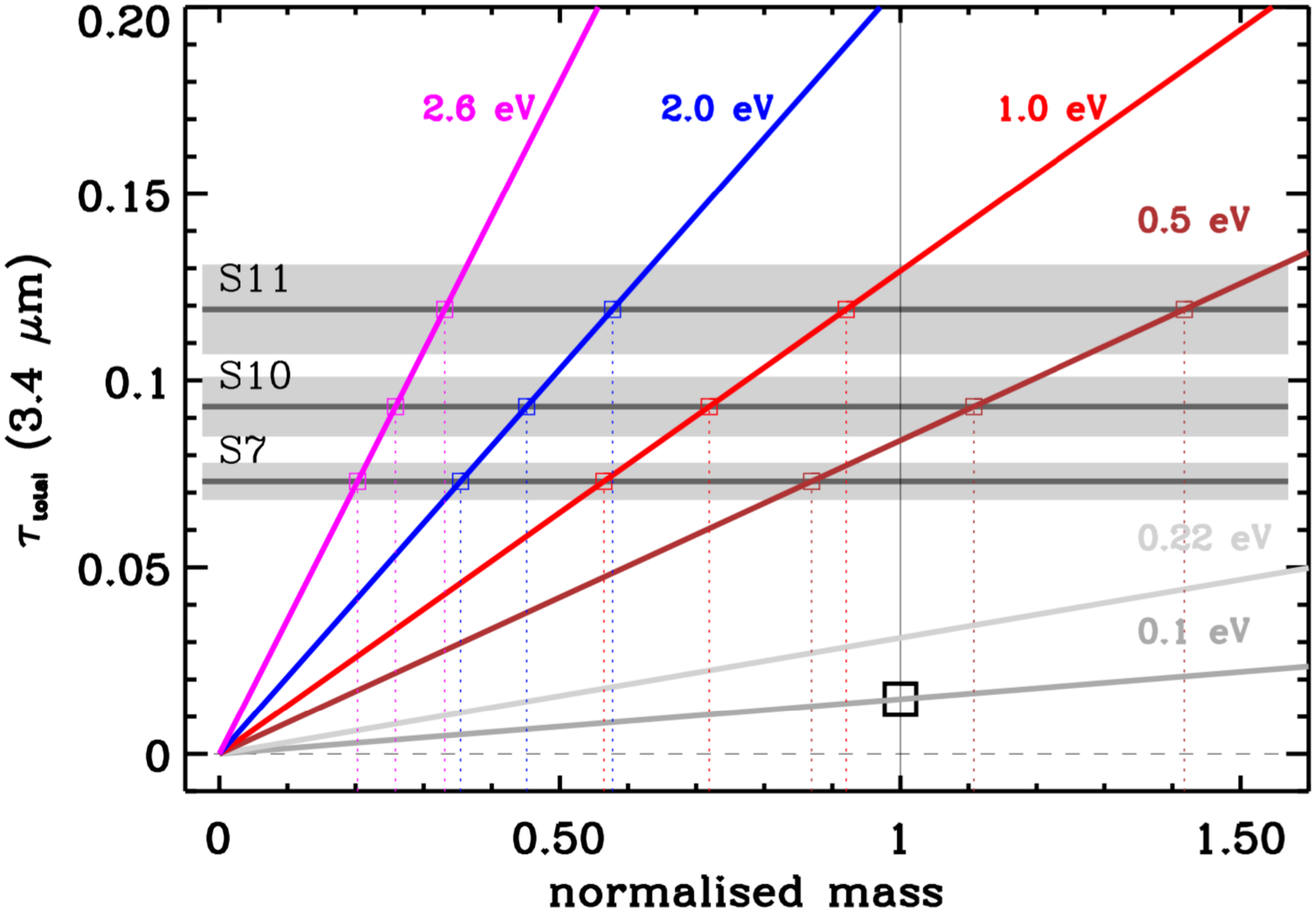}
\caption{The total optical depth at $3.4\,\mu$m, normalised to the dust model\cite{2013A&A...558A..62J,2014A&A...565L...9K}  mass (black square), as a function of the outer a-C(:H) mantle band gap (in eV). The horizontal bands show the observed optical depths and uncertainties along the S7, S10 and S11 lines of sight towards the IRAS~18511+0146 stellar cluster.\citep{2012A&A...537A..27G}}
\label{fig_3mic_tau}
\end{figure}

An unusually strong $3.4\,\mu$m interstellar absorption feature, on the long wavelength wing of the $3\,\mu$m water ice band,  is observed toward the IRAS~18511+0146 stellar cluster.\cite{2012A&A...537A..27G} 
In the diffuse ISM, where the carbonaceous dust is aromatic-rich and principally observed in emission, this feature is very weak.\cite{2013A&A...558A..62J}  
In the best fit to the diffuse ISM dust properties, using the new dust model (a-C(:H) mantles with $E_{\rm g} = 0.1$\,eV),\cite{2013A&A...558A..62J,2014A&A...565L...9K}  the optical depth in the $3.4\,\mu$m feature is only 0.015 (see Table \ref{table_tau} and Fig.~\ref{fig_3mic_tau}). 
We now explore whether and how the optical depths of the observed features \cite{2012A&A...537A..27G} could reasonably be explained by the addition of extra dust mass. 
In Fig.~\ref{fig_3mic_tau} and Table \ref{table_tau} the $3.4\,\mu$m feature total optical depth for the new model is shown as a function of the mantle material band gap and the dust mass normalised to that of the best-fit diffuse ISM model.\cite{2013A&A...558A..62J,2014A&A...565L...9K} Unlike the previous interpretation of these data the broad $3.47\,\mu$m feature was not removed prior to the optical depth determinations, which are therefore likely to be upper limits.  These data show that, for the standard model mantle composition, the dust mass must be increased by a factor of $\simeq 5-8$ in order to fit the $3.4\,\mu$m observational data towards IRAS~18511+0146. 
This is clearly unreasonable given that the quantity of accretable gas phase carbon could only increase the carbon dust mass by about a factor of two.\citep[{\it e.g.},][]{2012ApJ...760...36P,2013A&A...558A..62J,2015A&A...Ysard_etal}  
However, it appears likely that the additional mantling of CM grains, to form CMM grains, in the denser regions of the ISM must be of hydrogen-rich, wide band gap materials.\cite{Faraday_Disc_paper_2014,2015A&A...000A.000J,2015A&A...000A.000Y} 
On this basis, Fig.~\ref{fig_3mic_tau} and Table \ref{table_tau} indicate that the observed $3.4\,\mu$m optical depths could be explained by an increase in the dust mass by a factor of $1.2 - 1.9$, if the mantle is composed of an aliphatic-rich a-C:H material with $E_{\rm g} \gtrsim 1$\,eV. 
Recent modelling shows that in the transition from the diffuse to the denser ISM the dust evolves along the sequence CM $\rightarrow$  CMM $\rightarrow$ AMM\footnote{AMM are aggregates of CMM grains, which can then accrete ice mantles (I) to form AMMI aggregates} $\rightarrow$ AMMI.\cite{Faraday_Disc_paper_2014,2015A&A...579A..15K,2015A&A...000A.000J,2015A&A...000A.000Y} This evolutionary scenario was found to require the addition of aliphatic-rich a-C:H mantles, with an accompanying increase in the dust mass by a factor of $\simeq 1.6-1.8$,\cite{2015A&A...579A..15K} which then appears to be in reasonable agreement with the observational data.\cite{2012A&A...537A..27G}  
In this case an aliphatic $3.4\,\mu$m band ought to be polarised along these lines of sight because it coats all grains, including the large silicate grains, the $\sim 10\,\mu$m band of which is polarised. Currently the evidence for this along lines of sight towards the Galactic Centre indicates that the band is not polarised but these observations are difficult and the results still somewhat inconclusive.

% FIGURE 3

% TABLE *********************************************************
\begin{table}
\caption{The total optical depth at $3.4\,\mu$m, $\tau_{3.4\,\mu{\rm m}}$, as a function of the mantle material composition, as characterised by the optical band gap, $E_{\rm g}$. The entries in columns S7, S10 and S11 show the mass fraction, of a given band gap material, to be added to the diffuse ISM model dust\cite{2013A&A...558A..62J,2014A&A...565L...9K} in order to explain the observations toward the IRAS~18511+0146 stellar cluster \citep{2012A&A...537A..27G}.}
\begin{center}
\begin{tabular}{ccccc}
                           &                    &      &      &        \\[-0.35cm]
\hline
\hline
                          &                     &      &      &         \\[-0.25cm]
E$_{\rm g}$ [eV]   & $\tau_{3.4\mu{\rm m}}$ &  S7 & S10 &  S11   \\[0.05cm]
\hline
            &          &                             &        &         \\[-0.25cm]
    0.1   &     0.015       &   5.0     &   6.3    &  8.1  \\
    0.22  &    0.031       &   2.3     &   3.0    &   3.8 \\
    0.5   &     0.084       &   0.9      &   1.1    &  1.4  \\
    1.0   &     0.129       &   0.6      &   0.7    &  0.9  \\
    2.0   &     0.206       &   0.4      &   0.5    &  0.6  \\
    2.6   &     0.360       &   0.2      &   0.3     & 0.3  \\[0.05cm]
\hline
\hline
                          &                      &     &         &                   \\[-0.25cm]
\end{tabular}
\end{center}
\label{table_tau}
\end{table}

Within cloud interface regions and along molecular cloud lines of sight, where a-C:H mantle accretion occurs, the observed extinction curves appear to be "anomalous", in that they show steep UV and a weak and shifted (to $\sim 210$\,nm) or even absent UV bump, which is consistent with the extinction expected from a-C:H materials.\cite{2012A&A...542A..98J,2013A&A...558A..62J} 
Odd UV extinction curves do indeed appear to be the norm in these interface regions, {\it e.g.}, towards HD\,29647 (Taurus),\cite{1981MNRAS.196P..81W} HD\,62542,\cite{1993ApJ...408..573W} and also HD\,207198. 
In the case of HD\,29647 the dust extinction is characterised by $E(B-V) \sim 1$, $R_{\rm V} \sim 3.5$ and shows no evidence for H$_2$O$_{\rm (s)}$ ice mantles.\cite{1981MNRAS.196P..81W} 
The cometary globule sightline towards HD\,62542 is characterised by a high column density and low extinction ($A_{\rm V} \sim 1.20$ and $R_{\rm V} \sim 3.24$), typical of remnant molecular cloud material stripped by winds and UV radiation.\cite{1993ApJ...408..573W} 
Finally, the HD\,207198 line of sight has the largest carbon depletion, (C/H)$_{\rm dust} = 395\pm61$\,ppm, measured in the observed sample of 15 lines of sight.\cite{2012ApJ...760...36P} 
Such "anomalous" extinction curves are thought to be indicative of denser, molecular cloud material,\cite{1981MNRAS.196P..81W} which would be consistent with interstellar regions where a-C:H mantles have accreted onto all grain surfaces.\cite{2013A&A...558A..62J,2015A&A...579A..15K,2015A&A...000A.000J,2015A&A...000A.000Y}

%------------------------------------------------------------------
\subsection{C-shine}
\label{sect_precepts_Cshine}
%------------------------------------------------------------------ 

Early measurements of interstellar dust indicated a rather high albedo ($a \gtrsim 0.6$) at visible wavelengths.\cite{1936ApJ....83..162S,1937ApJ....85..194S,1941ApJ....93...70H,1968ApJ...152...59W,1970A&A.....8..273M,1970A&A.....9...53M} 
Later studies showed that in the near-infrared (NIR) the albedo in the J ($1.2\,\mu$m), H ($1.6\,\mu$m) and K ($2.2\,\mu$m) photometric bands is also high ($\sim 0.6-0.8$)\citep{1994ApJ...427..227W,1996A&A...309..570L} and that the NIR surface brightness of translucent and denser clouds is consistent with scattered radiation rather than dust emission.\cite{1996A&A...309..570L}  
Extinction mapping of the Perseus molecular cloud complex in the J, H and K bands, in regions with $A_{\rm V} < 30$\,mag revealed  cloudshine "emission" structures, which were interpreted as starlight scattering by dust in the clouds and assumed to be a measure of the dust mass distribution.\cite{2006ApJ...636L.105F} 
Related to cloudshine is an observed "emission" in the {\it Spitzer} IRAC 3.6 and 4.5\,$\mu$m bands, and absorption in the IRAC 5.8 and 8\,$\mu$m bands, termed "coreshine''.\cite{2010A&A...511A...9S}  
Cloudshine and coreshine (hereafter collectively called C-shine) have been interpreted in terms of IR scattering by big grains (radii $a_{\rm big} \simeq 1\,\mu$m) originating deeper within clouds and taken as evidence for significant dust growth there.  
However, more recent work points out that, while grain growth is indeed required, it is the particular nature of the outer a-C:H mantles (CMM grains) and the form of the aggregates of these grains (AMMI and AMMI) that are most likely at the origin of C-shine.
\cite{2015A&A...000A.000J,2015A&A...000A.000Y} Further, the coagulation of CMM grains into AMM and AMMI aggregates in this more recent work requires rather low levels of grain growth and can therefore occur on rather short time-scales with respect to cloud collapse and star formation. 
The growth of a-C:H mantles through the accretion of gas phase carbon onto dust in the outer reaches of molecular clouds may therefore provide a natural and self-consistent explanation for the the observed carbon depletion and extinction variations\cite{2012ApJ...760...36P,2013A&A...558A..62J} and C-shine.\cite{2015A&A...000A.000J,2015A&A...000A.000Y} As pointed out above,  this explanation of C-shine is also likely consistent with the absorption spectra along lines of sight that intersect denser interstellar matter.

%------------------------------------------------------------------
\subsection{Carbonaceous mantle evolution in the ISM}
\label{sect_mantle_evolution}
%------------------------------------------------------------------ 

Evidence seems to be growing that carbonaceous, a-C(:H), materials are an important component of ISM dust. 
\cite{1990QJRAS..31..567J,2008A&A...492..127S,2012A&A...540A...1J,2012A&A...540A...2J,2012A&A...542A..98J,2013A&A...555A..39J,2013A&A...558A..62J,Faraday_Disc_paper_2014,2014A&A...569A.119A,2015A&A...584A.123A}
Given that this material is more fragile than the silicate dust\cite{2008A&A...492..127S,2011A&A...530A..44J} it most probably exists as an abundant dust-mantling component in interstellar and solar system dust (see the above sections and Section \ref{sect_DustII}\ref{sect_DustII_organics} on organic nano-globules). The above-described THEMIS diffuse ISM dust model\cite{2013A&A...558A..62J,2014A&A...565L...9K} requires the presence of both aliphatic-rich (a-C:H) and aromatic-rich (a-C) carbonaceous dust components, with the former in grain cores protected by more resilient a-C mantle layers. 
A key question is then, how do such structures form, evolve and respond to their surroundings? Logically the more labile a-C:H component, which is sensitive to thermal and UV photo-processing in the ISM\cite{1990QJRAS..31..567J,2012A&A...540A...1J,2012A&A...540A...2J,2012A&A...542A..98J,2013A&A...555A..39J,Faraday_Disc_paper_2014,2014A&A...569A.119A,2015A&A...584A.123A} must be formed in regions protected from photolysis by the harsh interstellar UV-EUV radiation field, {\it i.e.}, in the dense UV-shielded regions of circumstellar shells, molecular clouds and pre-stellar nebul\ae. The subsequent and progressive exposure of the newly-formed a-C:H materials to photolysis, as theses dense regions disperse through the effects of circumstellar shell ejection or cloud disruption by star formation, will lead to the relatively gentle UV photo-processing as the local extinction diminishes. The result will be the de-hydrogenation of the outer a-C:H layers and its transformation into an aromatic a-C material, which is optically more opaque to UV photons and thus protects the underlying a-C:H. However, and based on experimental evidence,\cite{2008ApJ...682L.101M,2010ApJ...718..867M} it appears that the re-hydrogenation of a-C materials to a-C:H by atomic hydrogen addition at low temperatures ($T_{\rm gas} \simeq 80$\,K) also needs to be taken into account in the determination of the equilibrium composition and structure of composite a-C:H/a-C grains in the ISM. 

A fundamental question here concerns whether we even need an a-C:H core at all, in other words could a hollow shell or a solid a-C particle equally well explain the observations? 
Indeed, other than acting as a necessary accretion surface, the retention or preservation of a core within a shell structure is not essential because in the THEMIS CM dust model the a-C mantles reside on either a-C:H or amorphous silicate cores and both of these are dielectric materials with only weak absorption over significant portions of the spectral range. Such dielectric cores make little contribution to the shell material optical properties, over much of the UV-mm wavelength regime, and therefore play a relatively minor role with respect to a-C shell materials. However, the answer to the above-posed question would seem to be a qualified yes because, other than providing an accretion surface, some fraction of a-C:H grain materials are required in order to explain the shape of the unpolarised absorption band in the $3-4\,\mu$m region observed towards the Galactic Centre.\cite{1999ApJ...512..224A,2012A&A...540A...1J,2012A&A...542A..98J,2013A&A...558A..62J} 

% FIGURE 4
\begin{figure}[!h]
\centering\includegraphics[width=4.5in]{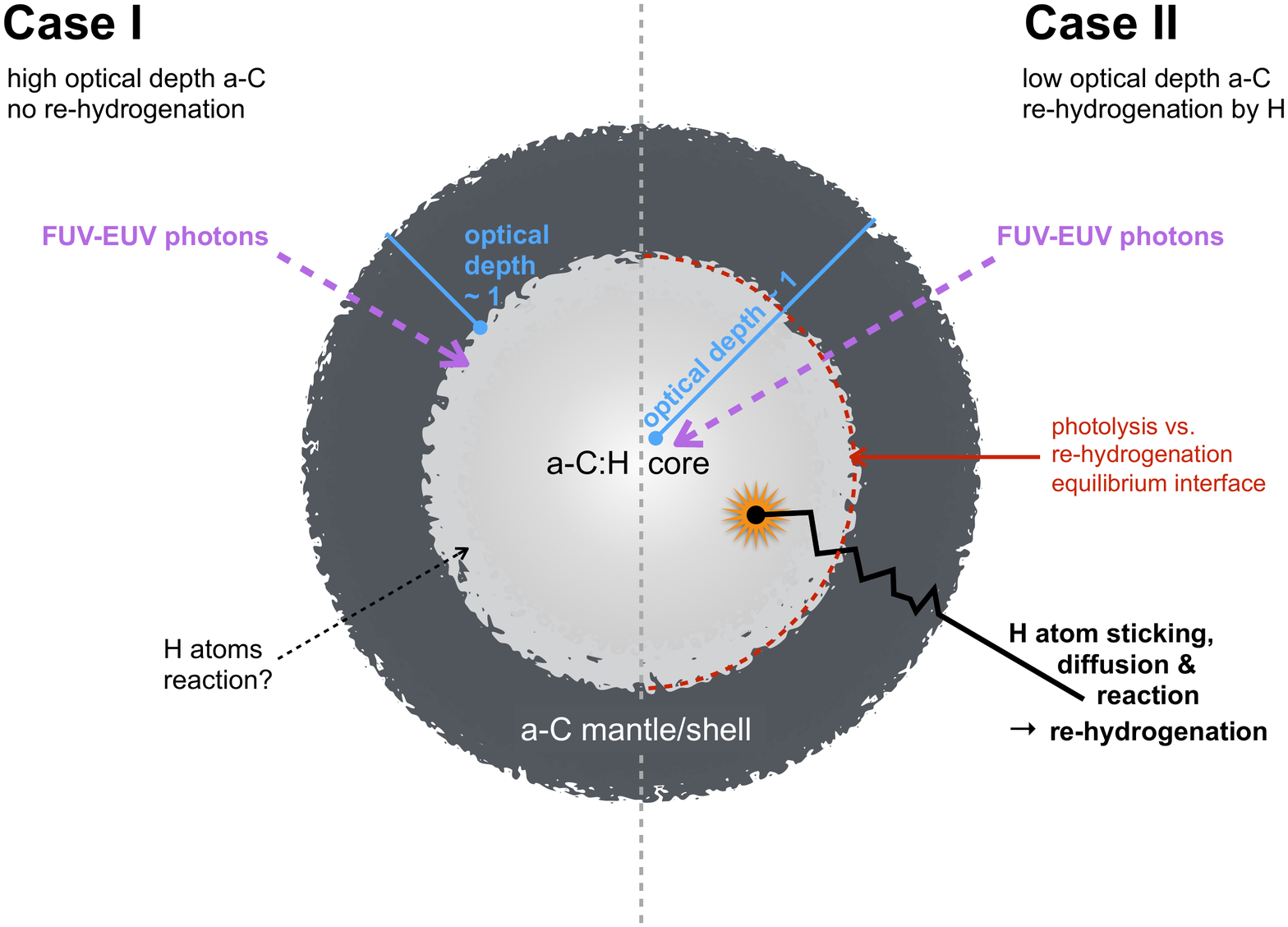}
\caption{The formation of core/mantle or shell structures in the ISM.}
\label{fig_CM_evolution}
\end{figure}

One can therefore imagine two scenarios (shown schematically in Fig.~\ref{fig_CM_evolution}) for the formation and maintenance of a-C:H/a-C core mantle grain structures in the ISM, depending on the relative rates of de-hydrogenative UV photolysis and re-hydrogenative H atom addition:

\noindent \underline{\bf Case I: high optical depth a-C and no re-hydrogenation:} Here the grain core/mantle structure is a direct result of the high optical depth of a-C at FUV-EUV wavelengths ($\equiv E_{\rm h\nu} \gtrsim 10$\,eV) and the intensity and hardness of the local interstellar radiation field (left portion of Fig.~\ref{fig_CM_evolution}). 
For both the modelled\cite{2012A&A...540A...2J} and laboratory measured\cite{2016arXiv160108037G} complex indices of refraction in the FUV-EUV wavelength regime the a-C(:H) optical depth is unity at a depth of $\simeq 25$\,nm and is relatively independent of the exact material composition, {\it i.e.}, be it a-C:H or a-C.\cite{2012A&A...540A...2J} 
Therefore, only the surface layers of the grain are de-hydrogenated to a-C, while the interior retains its original a-C(:H) composition, whatever that may be.\cite{2012A&A...540A...1J,2012A&A...540A...2J,2012A&A...542A..98J,2013A&A...558A..62J,Faraday_Disc_paper_2014} 
Thus, it is to be expected that carbonaceous grains will be completely UV photo-processed (to an aromatic-rich a-C material) to a depth of the order 25\,nm as has been hypothesised to be required for dust in the diffuse ISM.\cite{2012A&A...540A...2J,2013A&A...558A..62J,Faraday_Disc_paper_2014}

\noindent \underline{\bf Case II: low optical depth a-C and efficient re-hydrogenation:} Here FUV-EUV photolysis and a-C:H de-hydrogenation can act through a substantial fraction of the grain volume. This is compensated by an efficient re-hydrogenation by H atom sticking, diffusion and recombination, which converts a-C to a-C:H (right portion of Fig.~\ref{fig_CM_evolution}). In this case the core/mantle structure is likely to be very sensitive function of the finely-balanced equilibrium between photolysis and re-hydrogenation. The grain structure will therefore depend on both the intensity/hardness of the local interstellar radiation field and on the local environment density, {\it i.e.}, the H atom collision and incorporation rate.\cite{2008ApJ...682L.101M,2010ApJ...718..867M,2014A&A...569A.119A,2015A&A...584A.123A} 
In high radiation, low-density regions photolysis will have the upper hand and the grains will be predominantly aromatic-rich, while in low radiation, dense regions re-hydrogenation should win out and the grains will be transformed into aliphatic-rich particles. 

So, given the above comparison, is it possible to distinguish between these two cases through observationally-testable predictions? 
Also, in each case, how would the dust composition and structure be expected to evolve in the transition between diffuse and dense media and as interstellar matter cycles back-and-forth between these phases?  
Very generally, Case I ought to result in a rather stable dust configuration because the mantle/shell depth is determined only by the optical properties of a-C(:H) materials at FUV-EUV wavelengths. Further, Case I preserves/protects the core material, which therefore retains a history of its origin. 
In contrast, Case II depends on the equilibrium between the processes of FUV-EUV photolysis and re-hydrogenation by H atom addition. This perhaps ought to lead to wider dust variations throughout the ISM because its equilibrium composition/structure will depend on the local physical conditions (radiation field and gas density). Recent work indicates that this equilibrium is rather finely balanced, with a likely "switch-over" in diffuse/translucent regions with $A_{\rm V} \simeq 0.7$, which also appears to be where gas phase carbon accretion in the form of a-C:H mantles is taking place.\cite{Faraday_Disc_paper_2014,2015A&A...000A.000J,2015A&A...000A.000Y}
If Case II holds and re-hydrogenation were to dominate over photolysis in the diffuse/translucent ISM then this would lead to a significant fraction of aliphatic-rich material in the denser ISM and to large deviations in the shape of the UV bump and FUV rise in the extinction.\cite{2012A&A...542A..98J,2013A&A...558A..62J} While dust variations in the diffuse ISM are clearly observed they are nevertheless rather limited to relatively small variations in the dust opacity and the UV-FUV extinction.\cite{2007ApJ...663..320F,2015A&A...577A.110Y,2015A&A...580A.136F} 
However, lines of sight through the denser or higher column density ISM, where a strong FUV extinction associated with a weak UV bump, {\it e.g.}, towards HD\,29647,\cite{1981MNRAS.196P..81W} HD\,62542,\cite{1993ApJ...408..573W} and HD\,207198 seem to indicate significant carbon depletion from the gas.\cite{2012ApJ...760...36P} 
These extinction properties are characteristic of a-C:H materials which would likely have accreted as a-C:H mantles.\cite{2012ApJ...760...36P,2013A&A...558A..62J,2015A&A...000A.000J,2015A&A...000A.000Y} In conclusion, it is not perhaps yet possible to definitively determine whether ISM dust physics is on the side of Case I or Case II but the evidence does seem to lean slightly in favour of Case I in the diffuse ISM, where the dust is predominantly observed and characterised in emission.

%------------------------------------------------------------------
\section{Dust: evolutionary consequences}
\label{sect_DustI}
%------------------------------------------------------------------ 

If there is indeed, as seems highly likely, a significant transfer of matter back-and-forth between the gas and dust in transitional regions of the ISM, {\it i.e.}, matter at the interface between the diffuse and dense, molecular ISM, then there must be clear observational consequences. For instance, it is then to be expected that the more volatile element (O, C, N, S, \ldots) depletions vary with environment and the local dust chemistry. 
A specific example would be the effects of grain growth, via gas phase carbon accretion along with H atoms to form a-C:H mantles, contemporaneous with grain-grain coagulation, processes that are likely at the origin of the observed C-shine.\cite{2015A&A...000A.000J,2015A&A...000A.000Y}  
However, gas phase carbon and hydrogen atoms will not accrete alone but coincident with O, N, S, {\it etc.} depending on their charge state in the gas and that of the grains. 
For instance, carbon and sulphur accretion will be Coulomb-hindered in the diffuse ISM by the fact that both exist as ions   there and that the large grains are positively charged. However, in contrast to the larger grains, the carbonaceous nano-particles ($a \leqslant 3$\,nm) will be predominantly neutral\cite{2001ApJS..134..263W} and therefore capable of reaction with such ions as C$^+$, S$^+$, Si$^+$, Mg$^+$, Fe$^+$, {\it etc.}

In the following some key, promising and paradoxical aspects of dust evolution through surface chemistry are presented and explored within the encompassing framework of the THEMIS dust model. Along the way the consequences of a few perhaps unusual but very promising-looking dust chemistry scenarios are qualitatively explored.

%------------------------------------------------------------------
\subsection{Accretion anomalies}
\label{sect_DustI_accretion}
%------------------------------------------------------------------ 

In the ISM studies there are several outstanding issues that appear to be related to what is most likely an effect of anomalous accretion, by which it is meant that some process seems to be operating along lines other than the generally-accepted idea of passive accretion.  

The so-called oxygen depletion problem\cite{2009ApJ...700.1299J,2010ApJ...710.1009W} is a particular case in point. This arises from a detailed interpretation of observations\cite{2009ApJ...700.1299J} that clearly show that oxygen disappears from the gas at a rate faster than can be accounted for than by any obvious explanation, {\it i.e.}, by incorporation into a silicate/oxide dust or into an icy phase.\cite{2010ApJ...710.1009W} 
In order to explore this, if we were to assume that the oxygen depletion problem could be resolved by trapping it with another equally abundant and reactive species, then we are left with very little choice but to combine the oxygen missing from the gas phase with carbon and hydrogen in some solid phase. This would then implicate something like an oxygen-bearing carbonaceous material similar to cometary "organic" particles.\cite{2010ApJ...710.1009W}
In practice, and to solve the oxygen depletion problem, we would need to mop up $\sim 160$\,ppm of O\cite{2010ApJ...710.1009W} with only $\sim 160$\,ppm of C from the gas. However, we could also invoke the $\sim 120$\,ppm of C in the form of reactive or nascent a-C(:H) nano-particles that can also react with oxygen. With these two sinks we would, optimistically, have $\sim 280$\,ppm of C in a form that could react with oxygen atoms.
This would result in a material with [O]/[C] $\sim 160/280 = 0.6$ $\equiv$ C$_{1.7}$OH$_n$, or perhaps C$_{2}$OH$_n$ if the cosmic abundance of carbon is as high as 400\,ppm rather than 360\,ppm.\cite{2012ApJ...760...36P}   
In any event, and irrespective of the actual cosmic carbon abundance, any such $\approx$ C$_{2}$OH$_{(n < 6)}$ solid material ought to have observable oxygen-containing functional group IR signatures, {\it e.g.}, those of alcohol ($-$OH), carbonyl (C=O) and/or ether ($-$O$-$) bonds and should therefore be show its presence through the characteristic absorption bands of these functional groups.
However, there appears to be little if any observational evidence to support this scenario. 
Nevertheless, the observational evidence does show that there is a significant depletion of oxygen before the onset of ice mantle formation\cite{2010ApJ...710.1009W} and so it appears that oxygen must be strongly depleted into some as yet unknown solid form, which is not "ice like". 
However, it is hard to believe that all of the missing oxygen could be incorporated into an apparently "invisible" oxygen-rich organic carbonaceous solid phase. 
Perhaps a-C(:H) nano-particle epoxylation in the low density ISM\cite{ANT_RSOS_nanoparticles} might provide a part of the answer to this conundrum 
(see Sections \ref{sect_DustI}\ref{sect_DustI_epoxide} and \ref{sect_DustI_COseq}, and 
Sections \ref{sect_DustII}\ref{sect_DustII_organics} and \ref{sect_DustII_volatiles}).  

Several other interesting aspects of elemental depletion variations and dust accretion anomalies in the ISM have been explored within a global dust evolution framework, including "volatile" Si in photo-dissociation regions (PDRs), and likely N and S depletion sinks\cite{2013A&A...555A..39J}. 
In particular, this work pointed out that sulphur could be accreted into a-C:H mantles (at $\approx 10$\% with respect to carbon in dust) in the transition to the denser ISM, which would explain its disappearance from the gas. It seems most probable that sulphur must be trapped into a difficult to observe form. If indeed some fraction of sulphur were to be chemically-combined in  an a-C:H material then the  C$-$S and C=S  stretching bands in the $\simeq 15$ and $8-10\,\mu$m regions would be hidden by  the strong silicate absorption bands. 
Further, the association of sulphur with iron in FeS nano-inclusions in amorphous silicates is likely but it is not yet possible to constrain its abundance there from infrared observations.\cite{2014A&A...565L...9K}
In the ISM, and in contrast to sulphur, nitrogen does not appear to show a progressive depletion towards denser regions of the ISM but appears to maintain a low-level of depletion almost everywhere.\cite{2009ApJ...700.1299J}

%------------------------------------------------------------------
\subsection{Surface chemistry}
\label{sect_DustI_surfchem}
%------------------------------------------------------------------ 

In astronomical research grain surfaces (traditionally assumed to be of graphite or amorphous silicate) are generally treated as passive accreting surfaces taking no active part in interstellar chemistry.  
However, solid hydrocarbon a-C(:H) materials do without any question exercise a very active surface chemistry, especially their interaction with atomic O and N to form a-C:H:O:N:X hetero-atom doped substrates. 
Thus, this long-held view of passive interstellar grains surfaces is na\"ive, especially when it comes to the well-known nascent behaviour of nano-particles. Further,  it has been inferred that CH and OH are formed earlier than C$_2$, C$_3$, CN and CO in the ISM and that they therefore require precursor molecules in order to form them.\cite{2014IAUS..297..153W,2014IAUS..297...89Y} 
Thus, it would seem that nascent nano-particles could possibly be important catalysts for small radical and molecule formation in certain interstellar media.\cite{ANT_RSOS_nanoparticles} 

In the following sections some new and particularly promising scenarios for an active role for (nano-particle) grain surface-driven chemistry in the interstellar medium are explored in detail. 

% FIGURE 5
\begin{figure}[!h]
\centering\includegraphics[width=4.5in]{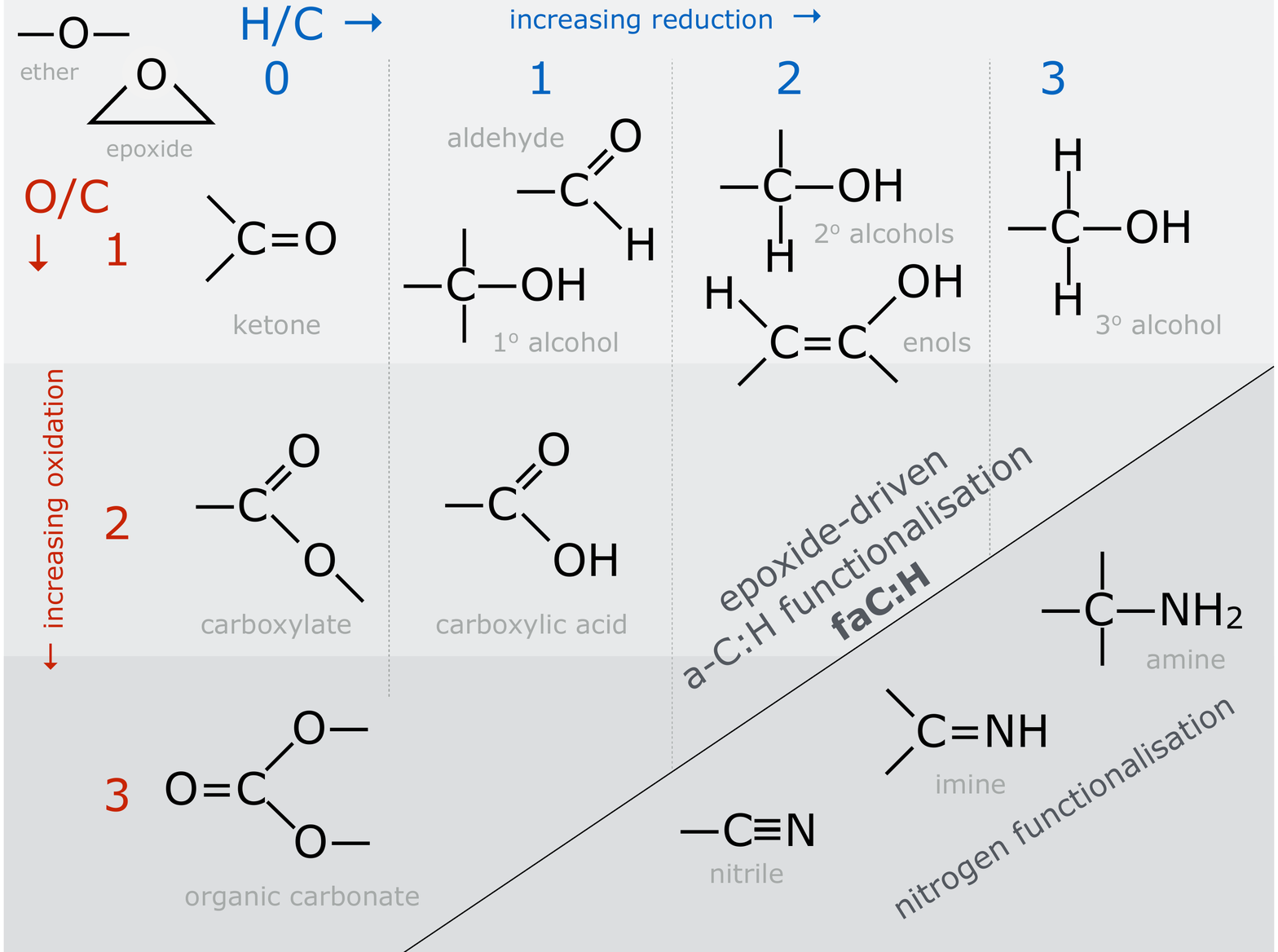}
\caption{Functional groups that can be incorporated into a-C:H and lead to its COH and CNH functionalisation.}
\label{fig_functional_groups}
\end{figure}

% FIGURE 6
\begin{figure}[!h]
\centering\includegraphics[width=5.5in]{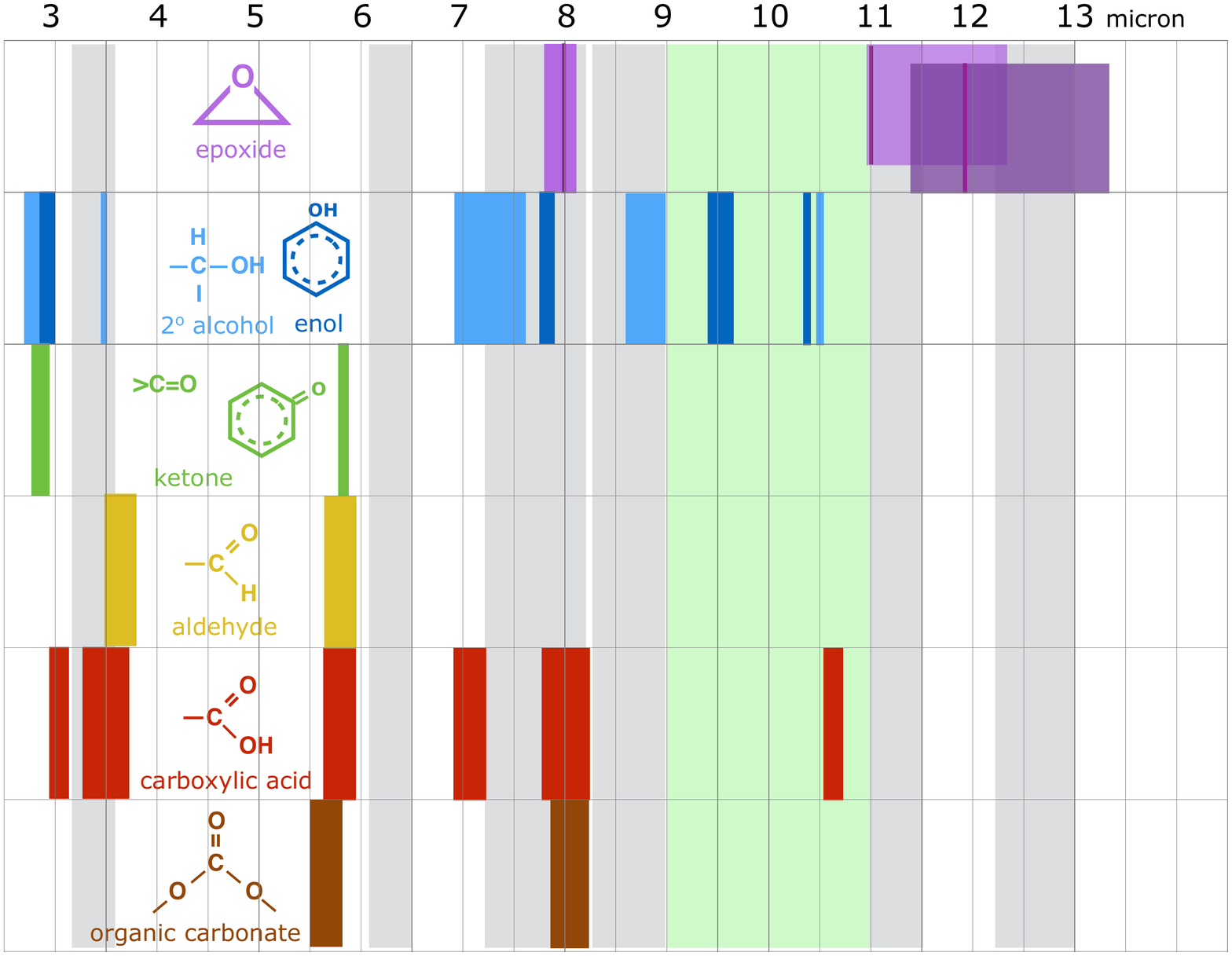}
\caption{Schematic view of the typical IR wavelength regions ({\it N.B.}, not the band widths) where the peaks of the given functional group absorption bands can be found. 
The wavelength in microns is shown on the upper scale. The grey bands indicate the approximate widths of the IR emission bands observed in the low-density, diffuse ISM and the green band the approximate width of the amorphous silicate $9.7\,\mu$m absorption band. 
For the epoxide the widely-variable positions of the two longer wavelength bands have been separated for clarity and the positions of the bands of a particular epoxide material are indicated by the thin darker lines.}
\label{fig_func_grp_IR_bands}
\end{figure}

% FIGURE 7
\begin{figure}[!h]
\centering\includegraphics[width=5.0in]{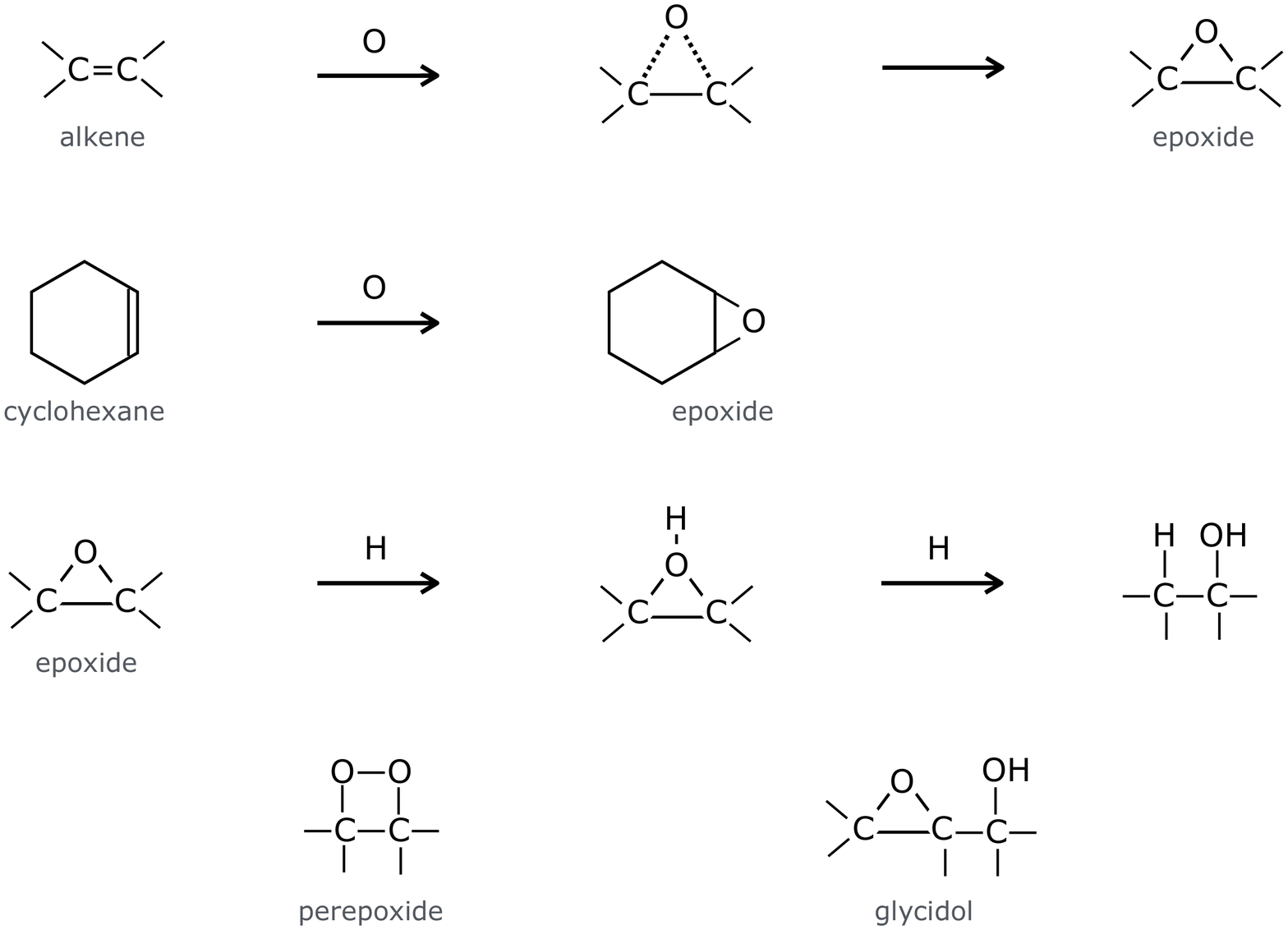}
\caption{Standard epoxide functional group formation, reaction and epoxide-related species. The sulphur analogues of these reactions are also likely to be viable.}
\label{fig_epoxide_form}
\end{figure}

% FIGURE 8
\begin{figure}[!h]
\centering\includegraphics[width=5.0in]{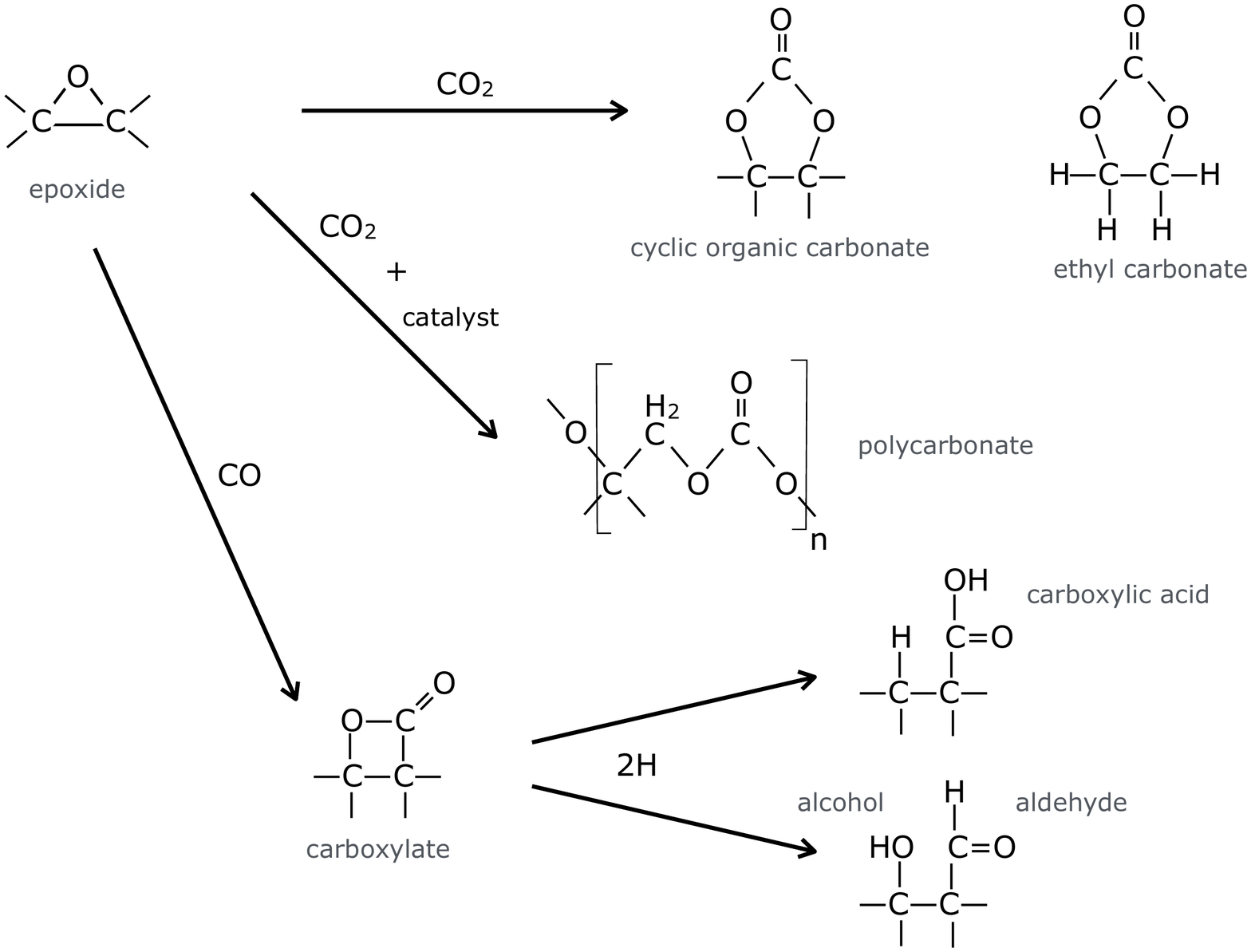}
\caption{Known epoxide reaction pathways with CO$_2$ and inferred reaction pathways with CO.}
\label{fig_epoxide_react}
\end{figure}

% FIGURE 9
\begin{figure}[!h]
\centering\includegraphics[width=5.5in]{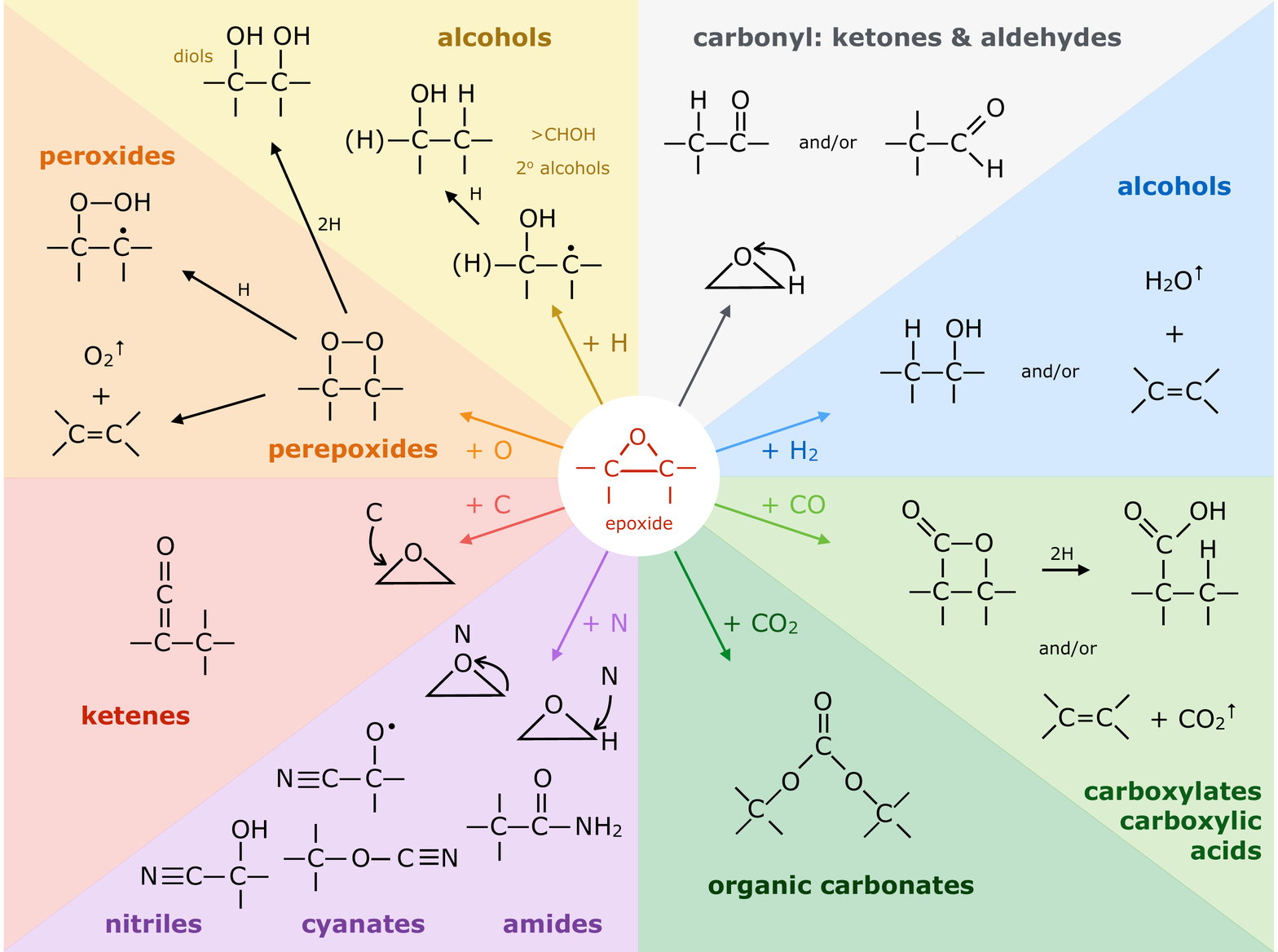}
\caption{A comprehensive set of epoxide reaction pathways.}
\label{fig_epoxide_pathways}
\end{figure}

% FIGURE 10
\begin{figure}[!h]
\centering\includegraphics[width=5.5in]{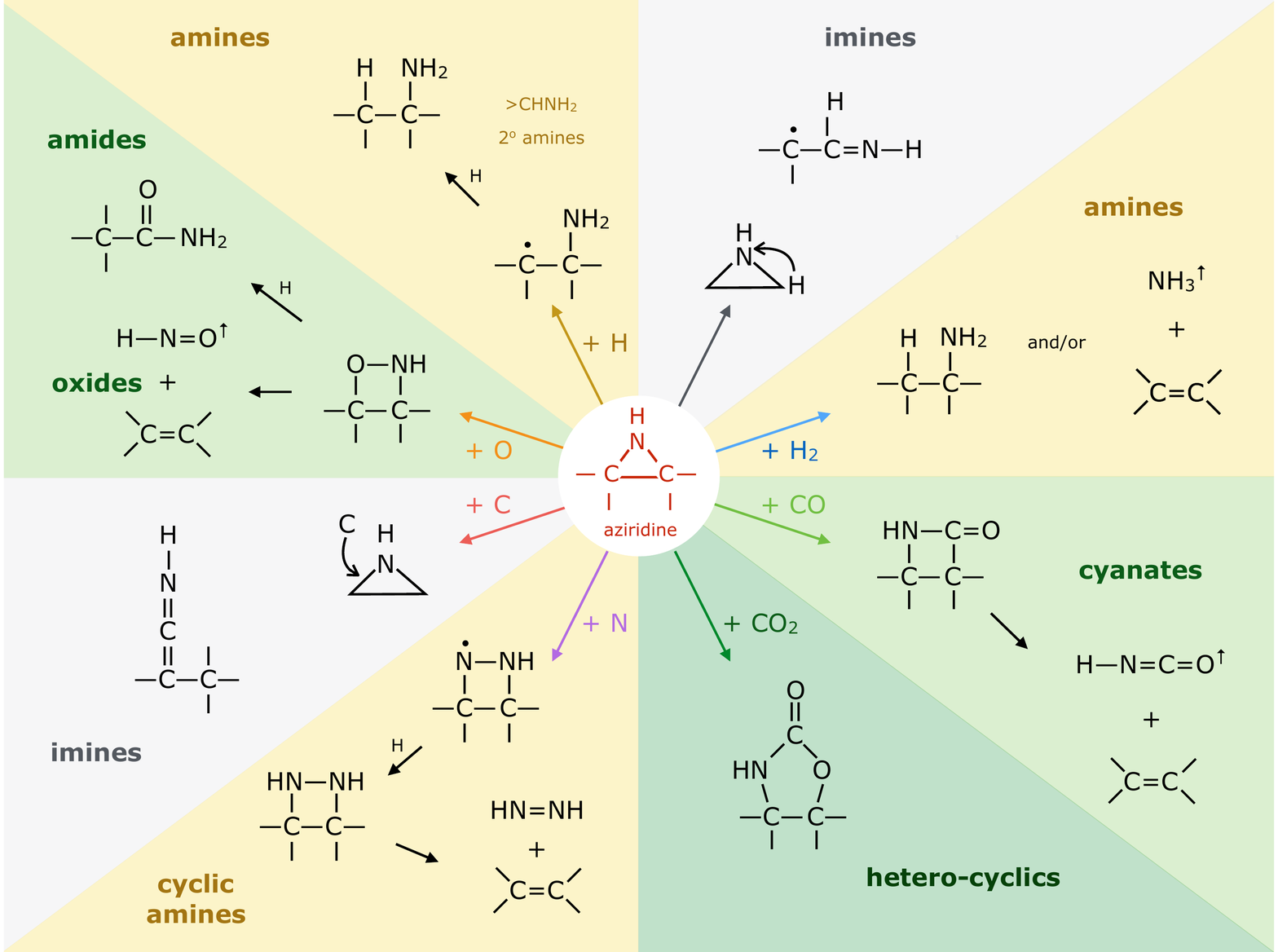}
\caption{A comprehensive set of aziridine reaction pathways.}
\label{fig_aziridine_pathways}
\end{figure}

% FIGURE 11
\begin{figure}[!h]
\centering\includegraphics[width=5.5in]{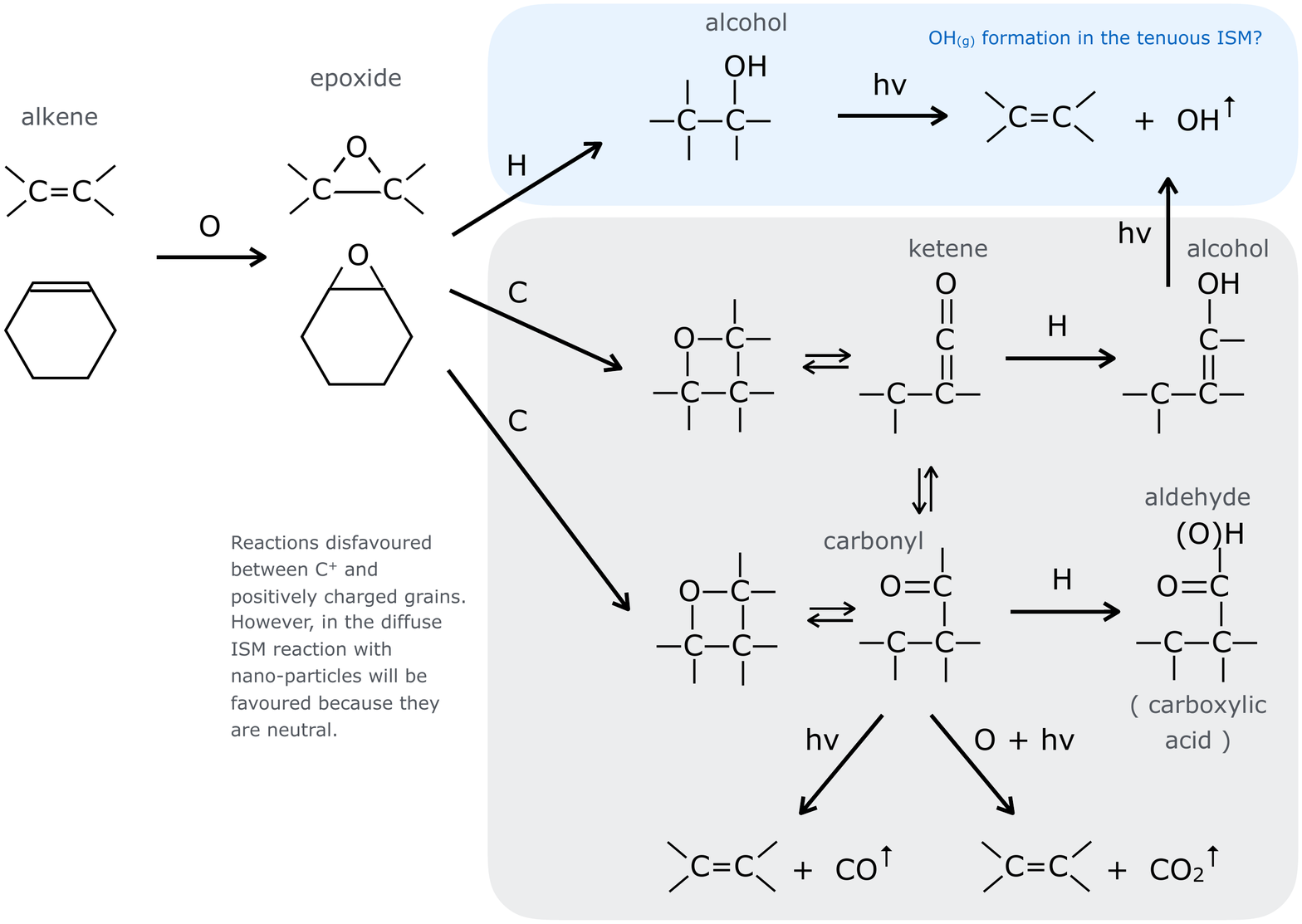}
\caption{Possible epoxide reaction pathways leading to the formation of OH, CO and CO$_2$ on carbonaceous nano-particles surfaces in the diffuse ISM. By analogy, reactions with episulphides on nano-particle surfaces would yield the sulphur analogue products SH$^\uparrow$, CS$^\uparrow$ and CS$_2^\uparrow$ released into the gas and $-$SH, $>$S, $>$C=S, $>$C=C=S, $>$C=C$<^{\rm SH}$, $>$C=C$\leqslant^{\rm SH}_{\rm O}$ and $>$C=C$\leqslant^{\rm OH}_{\rm S}$ surface functional groups.}
\label{fig_epoxide_OH_CO}
\end{figure}

%------------------------------------------------------------------
\subsection{Epoxides, aziridines and episulphides}
\label{sect_DustI_epoxide}
%------------------------------------------------------------------

In one of the companion papers\cite{ANT_RSOS_nanoparticles} it is shown that surface reactions on nascent a-C(:H) nano-particles could lead to some interesting chemistry and might provide a viable route to small polyatomic radicals whose presence in the tenuous ISM are not yet well-explained by current interstellar chemical models. In particular, it appears that reactions of gas phase atoms, principally O, N and S atoms, with olefinic and possibly also with aromatic C=C bonds\footnote{Possibly aided by the presence of magnesium or other metal hetero-atoms or cations within the structure.} could lead to the formation of labile three-fold ring species, {\it e.g.}, epoxide, $>$C$_-^{\rm O}$C$<$, aziridine, $>$C$_-^{\rm N}$C$<$ and episulphide, $>$C$_-^{\rm S}$C$<$, functional groups (where C$_-^{\rm X}$C represents a $\triangle$ three-atom ring structure), which decompose under the effects of UV irradiation to yield gas phase OH, NH and SH. The episulphide ring  is significantly less stable that the others and so would be a particularly reactive species in the ISM. 
Such an epoxide reaction route\cite{ANT_RSOS_nanoparticles} does appear to be compatible with the observed relative abundance of OH radicals in the cirrus clouds of the North Celestial Loop.\cite{2010MNRAS.407.2645B} 
Rather intriguingly, some cometary organics have been found to be soluble when embedded in epoxy. This phase is compositionally similar to, but distinct from, epoxy and spectroscopically slightly different from the embedding epoxy in that it consists of labile aliphatics.\cite{2009LPI....40.2260D} 
Thus, it appears that pre-solar organics may have retained a fossil chemical signature of their epoxylated olefinic-rich and aliphatic-rich components, which formed during their evolution in the low-density ISM, before the onset of cloud collapse and molecular cloud formation that led to the formation of the pre-solar nebula. 

Generally, under the conditions typical of the tenuous ISM, oxygen is atomic but both carbon and sulphur are singly-ionised (C$^+$ and S$^+$). Thus, the reaction of C$^+$ and S$^+$ ions with positively-charged grains would be inhibited. 
However, in the diffuse ISM the carbonaceous nano-particles ($a \lesssim 3$\,nm), which carry the bulk of the react-able or nascent interstellar grain surface,\cite{2013A&A...558A..62J} are predominantly electrically-neutral.\cite{2001ApJS..134..263W}
Thus, both C$^+$ and S$^+$ ions can also interact with nascent nano-particle surfaces and insert into olefinic or aromatic C$=$C bonds to form reactive, three-atom, strained-ring structures. 
Therefore, and in addition to the simple diatomic hydrides mentioned above, reactive surface epoxide, aziridine and episulphides could also yield interesting species such as C$\equiv$O, C=C=O, c-C$_2$O, C$\equiv$N, C$\equiv$S, C=C=S, c-C$_2$O, H-C=O, H-C=C=O, H-C$\equiv$N, H-N$\equiv$C, H-C=N-H, H-C=S, H-C=C=S \dots  \ 
In addition, C$^+$ insertion into surface C=C bonds, to form cyclopropene-like rings, could 
subsequently lead to the formation of species such as H-C=C=N-H, l-C$_3$H (C=C=C-H) and c-C$_3$H and their liberation into the gas in low-density regions. 

The incorporation of O (and probably also CO, see following section) into epoxide-actived sites on and within interstellar a-C(:H) grains may perhaps have some bearing on the so-called oxygen depletion problem\cite{2009ApJ...700.1299J,2010ApJ...710.1009W} (see above). Similarly, the incorporation of sulphur into reactive episulphides, which are more reactive than epoxide and aziridine species and are also known to polymerise, may provide a mechanism for depleting sulphur in the ISM. However, a good determination of the sulphur depletion in the ISM is rather troublesome because of the limited number of useful observations but it does, nevertheless, appear that it is depleted to some extent.\cite{2009ApJ...700.1299J} Hence, the question of sulphur depletion from the gas must remain open until such time as observations provide more stringent constraints. 

Epoxide functional groups (and by inference aziridine and episulphide groups) on grain surfaces may therefore play a key role in interstellar chemistry at the interface between the tenuous and dense ISM, particularly under the prevailing conditions at the onset of ice mantle formation. 
The formation of epoxide groups in the laboratory is achieved by the reaction between peroxide $-$O$-$O$-$ and alkene  $>$C=C$<$ species, including six-fold alkene-containing rings such as cyclohexene, through peroxide-released O atom insertion into the alkene double bond, {\it e.g.}, 
$>$C=C$<$  + O $\rightarrow$ $>$C$_-^{\rm O}$C$<$ (see Fig.~\ref{fig_epoxide_form}). 
The resulting epoxide group is a cyclic ether with a highly-strained three-fold ring, which is particularly reactive and sensitive to light, a basic requisite of epoxy-based adhesives. 
For example, the epoxide group is known to react with hydrogen to form alcohols, under illumination, as shown by the lower reaction in Fig.~\ref{fig_epoxide_form}.
Thus, epoxide formation leads to the oxidation of alkenes, including the formation of carbonyl bonds ($>$C=O). 
Indeed, the chemical synthesis of organic polycarbonates, through the reaction of carbon dioxide with epoxides, can be used to sequester carbon dioxide.  This synthesis route readily forms cyclic organic carbonates but in the presence of an organic ligand, bimetallic catalyst produces a high yield ($\leq 50$\%) of organic polycarbonates. 
Given that epoxides react readily with CO$_2$ to form organic carbonates, it would seem that they must also react with carbon monoxide to form carbonyl bonds in ketones, carboxylates and carboxylic acid functional groups, in addition to organic (poly)carbonates. 
The known and some probable reaction pathways for epoxide reactions leading to a variety of alcohol and carbonyl-containing function groups, {\it e.g.}, ketones, aldehydes, carboxylic acids, carboxylates and organic (poly)carbonates, are shown in Fig.~\ref{fig_epoxide_react}. 
It is also likely that the sulphur analogues of the reactions shown in Fig.~\ref{fig_epoxide_react} may also provide viable reaction pathways in the ISM. 

A comprehensive set of likely grain-component epoxide (episulphide) and aziridine reaction pathways with abundant interstellar atoms is shown in Figs.~\ref{fig_epoxide_pathways} and \ref{fig_aziridine_pathways}, respectively. From these figures it is apparent that the possible surface epoxide group reactions with gas phase species are more numerous and more varied than the equivalent aziridine reactions, as indicated by the need for a broader colour scheme in the epoxide figure (Fig.~\ref{fig_epoxide_pathways}). 
Clearly, the full astro-chemical implications of these reaction pathways will require their incorporation into a full chemical network model. However, given that most of the exact channels, branching ratios and rates are unknown, this would seem to be a rather premature exercise at the moment. 

In the ISM it is possible that nitrogen and carbon atoms can also be involved in these and related chemical pathways. However, as proposed in the companion papers,\cite{ANT_RSOS_nanoparticles,ANT_RSOS_topdown} it is likely that nitrogen atoms once incorporated into carbonaceous dust will likely form rather stable heterocyclic pentagonal rings associated with aromatics. 
In the tenuous ISM, where carbon is ionised (C$^+$) and the larger grains are positively charged, surface reactions with carbonaceous grains will likely be suppressed. However, the predominant grain surface-carrying nano-particles are mostly neutral in these same regions and so the reaction of C$^+$ (and by analogy S$^+$) at these nascent grain surfaces is not inhibited. Nevertheless, it is likely that most of the accreted carbon atom/ions would incorporate into the grain structure, leading to grain growth. 
Hence, the most likely product of carbonaceous nano-particle nascence is the formation of OH radicals,\cite{ANT_RSOS_nanoparticles} with NH (SH) less abundant by a factor of at least 5 (30), given the cosmic abundances of these elements. 
The possible nano-particle surface reactions with O (S) and C atoms are summarised in Fig.~\ref{fig_epoxide_OH_CO}. 

The molecule ethylene oxide (c-C$_2$H$_4$O, {\it i.e.}, $^{\rm H}_{\rm H}>$C$_-^{\rm O}$C$<^{\rm H}_{\rm H}$) has been detected towards the Galactic centre, with an abundance relative to hydrogen of $X$[c-C$_2$H$_4$O] $= 6 \times 10^{-11}$, which is $\geq 200$ times larger than could be explained by the available chemical models; suggesting that grain surface chemistry may play a role in its formation.\cite{1997ApJ...489..753D}
Ethylene oxide has also been detected in a number of hot molecular cores, with $X$[c-C$_2$H$_4$O] $= (2-6) \times 10^{-10}$, where its isomer acetaldehyde (CH$_3$CHO, {\it i.e.}, CH$_3-$C$\leqslant^{\rm H}_{\rm O}$) is detected at an abundance $2.6-8.5$ times that of ethylene oxide and where methanol (CH$_3$OH), dimethyl ether (CH$_3$OCH$_3$), ethanol (C$_2$H$_5$OH) and methanal (HCOOH) are also observed.\cite{1998A&A...337..275N} 
It has been shown that gas phase formation routes to acetaldehyde, and its derivatives, appear to be inefficient\cite{2004AdSpR..33...23C} and so a grain surface formation route to epoxides, aldehydes and other carbonyl-containing species such as ketones and carboxylic acids does indeed look attractive. 

Laboratory experiments on epoxide formation show that atomic oxygen reacts with both ethene (H$_2$C=CH$_2$) and propene (H$_2$C=CHCH$_3$), at $12-90$\,K, to form epoxides (on an HOPG graphitic surface) with lower energy barriers than for the equivalent reactions in the gas phase.\cite{2011ApJ...741..121W} 
For both of the studied alkenes the reaction yields peak at $T \sim 40-80$\,K, drop off rapidly for $T > 60$\,K but are still rather efficient over the temperature range $0-90$\,K. 
For ethene the peak reaction efficiency is $\sim 50$\% for $T \simeq 30-50$\,K but only $\sim 5$\% at $20$\,K. 
However, the reaction with propene is more efficient at these same temperatures, which presumably implies that it will also be very efficient for larger carbon-carbon double bond containing structures such as olefinic/aromatic-rich interstellar nano-particles, {\it e.g.},  "\ldots {\em processing of longer chain alkenes by oxygen atoms should be highly efficient} \ldots".\cite{2011ApJ...741..121W}
Experimental studies therefore seemingly favour grain surface routes to epoxide formation within an ISM context. 
Interestingly, and other than epoxide formation, these experiments also yielded small quantities of acetaldehyde (CH$_3$CHO) but no vinyl alcohol (H$_2$C=C$<^{\rm OH}_{\rm H}$).\cite{2011ApJ...741..121W}
Thus, "\ldots {\em at temperatures mimicking those in interstellar clouds, oxygen atoms can readily add to carbon-carbon double bonds to produce epoxide rings.}"\cite{2011ApJ...741..121W} 
 
Clearly, if epoxides do play a role in interstellar chemistry in low density regions they ought to somehow reveal their presence there. 
The detection of epoxide groups, $>$C$_-^{\rm O}$C$<$, present in interstellar dust is, in principle, possible through the observation of their characteristic bands that fall within the wavelength regions $\simeq$\,7.8$-$8.1, {\bf 10.9$-$12.3} and {\bf 11.4$-$13.3}\,$\mu$m (with the stronger bands indicated in bold face type). However, in the low density ISM all of these bands lie within wavelength regions that will be dominated by the aromatic-rich material IR emission bands and the epoxide bands will therefore be masked by other more abundant IR dust features (see Fig.~\ref{fig_func_grp_IR_bands}). Indeed, it has been shown that the IR spectroscopy of small carbonaceous molecules, such as ethylene oxide (c-C$_2$H$_4$O) and cyclopropenylidene (c-C$_3$H$_2$), shows a remarkably-interesting correspondence with most of the so-called aromatic emission bands.\cite{2009ApJ...704..226B} 
Further, it has been proposed that ethylene oxide is formed and retained on carbonaceous grain surfaces, precluding the radio detection of its rotational lines.\cite{2009ApJ...704..226B} 
The viability of ethylene oxide formation on grain surfaces appears to be further supported by modelling.\cite{Occhiogrosso:2012vl,Occhiogrosso:2014it} Fundamentally related to this work but clearly distinct from the idea of molecular ethylene oxide, it is proposed here that epoxide structures form on (nano-)particle surfaces in the low density ISM and that they are an intrinsic part of the contiguous grain structure in these diffuse regions. 
There they invest the grains with a nascence that can drive an interesting and diverse suite of chemical reactions. 
Hence, the direct detection of epoxide species is inconveniently difficult. 
In denser regions, where the grains are ice-mantled, ethene is not detected within the ices and so it would seem that epoxide functional groups are probably not associated with the ices, where they would be expected to be highly reactive. Instead, they must have existed, as an integral component of the underlying grain surfaces rather than as a discrete molecular species within volatile ices. Thus, they would have reacted with abundant gas phase species, {\it i.e.}, O and C, prior to ice formation, and yielded surface carbonyl-rich surface functional groups such as ketones, aldehydes, carboxylic acids and carboxylates, as well as secondary alcohols (see Fig.~\ref{fig_epoxide_pathways}) and their sulphur analogues.  Similarly, reactions with aziridine-type surface functional groups would yield lower abundance amine, imine, amide and cyanate surface groups (see Fig.~\ref{fig_aziridine_pathways}).

Thus, it is possible that epoxides are present in dust in the diffuse ISM and that their signatures have not yet been fully recognised as such. For example, the ISO SWS spectra of interface regions, {\it i.e.}, regions where fresh molecular cloud dust material is newly exposed to stellar radiation, reveal rather broad component sub-bands in the $\simeq 7.6-8.1$ and $\simeq 10.5-12.3\,\mu$m wavelength regions that have been attributed to the suite of PAH bands\cite{2001A&A...372..981V} but which could perhaps equally well be attributed to epoxide species within the dust (epoxide-containing, aromatic-rich dust). 
While the attribution of the majority of the aromatic emission bands to ethylene oxide and cyclopropenylidene molecules  might perhaps be something of an over-enthusiastic interpretation of the data, there is nevertheless a sufficiently-intriguing correspondence between the epoxide ring modes\footnote{The 7.8$-$8.1, {\bf 10.9$-$12.3} and {\bf 11.4$-$13.3}\,$\mu$m bands are attributable to epoxide ring stretch ("breathing"), asymmetric and symmetric ring deformation modes, respectively.} and some of the emission bands. It is therefore not entirely impossible that surface epoxide groups on nano-particles could make a contribution to some of the fundamental IR emission bands. 
In particular, epoxide bands in the $\simeq 8$ and 11$-13\,\mu$m wavelength regions could contribute significantly to the 7.8, 11.0, 11.3 and $12.7\,\mu$m interstellar "aromatic" emission bands and might even be the origin of some of them.

%------------------------------------------------------------------
\subsection{Carbonyl formation and CO sequestration}
\label{sect_DustI_COseq}
%------------------------------------------------------------------ 

Three interesting conundrums in interstellar chemistry are, 
firstly why does OH appear to form before CO? 
Secondly, why does CO appear to be under-abundant with respect to molecular hydrogen in transitional interstellar clouds, {\it i.e.}, what is the origin of the so-called "CO dark gas"?  
Thirdly, why does oxygen disappear from the gas, in the transition to denser interstellar regions, at a rate that cannot be accounted for by its incorporation into refractory or icy solids? 
The latter question is the the so-called "oxygen depletion" problem. 
Given that these problems appear occur under similar conditions it is perhaps not an unreasonable hypothesis to assume that the three are somehow related. 
As proposed above, and in the companion paper,\cite{ANT_RSOS_nanoparticles} the chemistry in tenuous interstellar regions is likely driven in no small part by nascent nano-particle surface chemistry. 
Given that, other than hydrogen atoms, oxygen atoms are likely to be the most abundant and also the most reactive gas phase species they will be the major driver of nano-particle surface-chemistry in transition regions. 
The carbonaceous dust-driven formation of CO molecules in the tenuous ISM will be disfavoured by the obvious chemical affinity for accreted carbon to be retained on/within carbonaceous grains. Also, any CO that is formed, via either gas- or dust-driven reactions will likely be preferentially sequestered onto the surfaces of nano-particles where it would become an intrinsic part of the grain surface structure within carbonyl functional groups. Thus, CO formation and retention in the gas would appear to be discouraged in the low density ISM. 
This scenario would appear to provide a coherent explanation for the CO dark gas problem because in tenuous regions CO formed on and retained on dust and gas phase CO, of what ever origin, will be sequestered onto/into dust in the form of carbonyl-containing functional groups (ketone, aldehyde and carboxylic acid, \ldots). In contrast, gas phase molecular hydrogen will not be accreted onto dust and will remain in the gas. This scenario can then explain the existence of a "CO dark" interstellar medium phase (with H$_2$ but little or no CO) and the presence of OH radicals within this same gas,  thus providing a self-consistent connection between these two problems.
A clear and self-consistent solution to the oxygen depletion problem would appear to be natural consequence of O atom and CO molecule sequestration from the interstellar gas via reactive nano-particle surface chemistry (nascence), which would lead to the formation of grain-surface oxygen-rich functional groups, {\it e.g.}, ketone, aldehyde, carboxylic acid, carboxylate and organic carbonate groups. 
Clearly, such species ought then to be detectable in the tenuous ISM in regions that have not yet or are just beginning to accrete ice mantles, {\it i.e.}, onset ice mantle regions. 
However, from Fig.~\ref{fig_func_grp_IR_bands} it is evident that practically all of the characteristic IR bands of these O-rich functional groups fall within or close to the positions of strong dust emission bands in the tenuous ISM. Thus, it would appear that it is conveniently rather difficult to test this hypothesis with existing data as it requires a sensitive search for weak and broad IR absorption bands in the low density ISM, in regions where the dust emission bands are weak or absent, {\it i.e.}, most likely in the outer reaches of molecular clouds. 

The above scenario for the anhydrous formation of carbonyls and carbonates, therefore implies that the presence of carbonates in interstellar and/or pre-solar grains does not necessarily require the presence of liquid water for their formation.  

Atomic oxygen accretion, and also CO molecule accretion/sequestration from the gas, will drive surface chemistry and lead to the formation of alcohol and carbonyl-containing (ketone, aldehyde, carboxyl, carbonate, \ldots) functional groups  (see Figs.~\ref{fig_functional_groups} to \ref{fig_epoxide_OH_CO}).
These accretion-driven reactions probably occur prior to or at the onset of ice mantle accretion and could provide an explanation for the carbonyl band observed in the ISM of the nuclear region of the Seyfert 2 galaxy NGC\,1068.\cite{2004A&A...423..549D}

Further, the seemingly rather intimate link between interstellar nano-particle nascence and chemistry, the spatial correlation between highly-excited $^{12}$CO, $^{13}$CO and warm dust in the dense and filamentary structures at the edge of PDRs\cite{2014A&A...569A.109K} is then perhaps not so surprising.

%------------------------------------------------------------------
\subsection{Episulphide formation and sulphur sequestration}
\label{sect_DustI_Sseq}
%------------------------------------------------------------------ 

By a somewhat analogous pathway to the sequestration of CO from the gas and into dust, sulphur ions could react with nano-particle surfaces in the diffuse ISM to form episulphide groups that are part of the contiguous grain structure. 
However, given that episulphides are more reactive than epoxides and that they tend to polymerise, they are likely to react to form species that are retained on the surface. This could closely follow analogues of the above pathways proposed for CO sequestration in carbonyls and organic carbonates. However, the sulphur in organo-sulphur compounds preferentially takes on a stable bridging role in a large number of both chain-like and ring molecules, {\it e.g.}, as in thioethers, $-$S$-$, disulphides, $-$S$-$S$-$, singly/doubly S-substituted pentagonal rings and as a bridge between aromatic rings as in phenoxathiin, which consists of two benzene rings connected by a sixfold ring with $-$S$-$ and $-$O$-$bridges ({\it i.e.}, {\large $\varhexagon$}$^{\rm O}_{\rm \hspace{0.03cm}S}$\hspace{-0.02cm}{\large $\varhexagon$}). 
Thus, a more likely scenario for sulphur sequestration and depletion in the ISM would be that sulphur ions are first trapped or incorporated into reactive episulphide functional groups on nano-particle surfaces. These episulphide groups could then react with H atoms from the gas to form surface thiols $-$SH or release SH into the gas. Although, more likely they will react with adjacent carbon atoms in the particle surface and/or with incident gas phase heavy atoms, predominantly O, C$^+$, N and S, which would likely tend to open up the  three-fold episulphide ring to form larger and less-strained bridging structures. 

Thus, it appears that nano-particle surface reactions with S$^+$ ions could provide a viable explanation for sulphur depletion from the gas. Such a scenario would be self-consistent with the other nano-particle surface reaction pathways proposed here and therefore appears worthy of a more quantitative evaluation once the appropriate reactivities and their rates have been experimentally-determined.

%------------------------------------------------------------------
\subsection{Silicon in PDRs and the origin of SiO in shocks}
\label{sect_DustI_Si_SiO}
%------------------------------------------------------------------ 

The dust formed around oxygen-rich evolved stars incorporates essentially all of the available silicon into amorphous and a small fraction of crystalline silicates. However, in the cold diffuse ISM $\sim 10$\% of silicon  is observed in the gas phase and this fraction rises to $\sim 50$\% in warm galactic halo clouds.\cite{1996ARA&A..34..279S}
Further, anything ranging from 10$-$50\% of silicon is to be found in the gas phase in galactic PDRs,\cite{2000A&A...356..705R,2003A&A...412..199O,2006ApJ...640..383O} reflection nebul\ae,\cite{2000A&A...354.1053F} and HII regions.\cite{1993ApJ...413..237C} Such high fractions for gas phase silicon appear to be the same as those found in violently-shocked regions of the ISM, {\it i.e.}, 10$-$40\% of Si in the gas.\cite{2002ApJ...579..304W,2006A&A...456..189P,2009SSRv..143..311S,2012A&A...548L...2C} This poses something of a conundrum because PDRs, reflection nebul\ae\ and HII regions are comparatively-benign environments for Si-containing dust and are therefore not expected to be as destructive as proto-stellar jets and supernova-generated shocks. While all of the Si abundance determinations for shocked regions rely on distinctly-different NIR, optical and UV and mm line measurements (made with ESO 3.6m telescope, NTT, HST and IRAM instruments), it is perhaps somewhat curious that the unexpectedly-high PDR, reflection nebula and HII region determinations all rely on the observation of a single Si$^+$ line at 35$\,\mu$m (made with KAO and ISO instruments). Thus, if the oscillator strength for this line is erroneous, then all of these benign region Si abundance measurements will be in error by the same mis-measure. 

If, however, a high gas phase abundance of Si in relatively benign environments is supported by the data, then this could be consistent with silicon being incorporated as a dopant into accreting a-C:H mantles in the denser ISM and then later being released back into the gas phase via dust photo-processing in the PDRs associated with young stars.\cite{2013A&A...555A..39J} Here the silicon will exist as Si$^+$ but in high-temperature shocked regions the released Si, either neutral or ionised, will react with gas phase oxygen to form the SiO that is used as a ubiquitous shock-tracer. It is also possible that some SiO could be released directly from eroded a-C(:H) grain mantles where a fraction of it could be bonded to O atoms as a result of epoxide-driven or other surface oxygen species reactions. 
It is interesting to note that in shocked regions, where a significant fraction (10$-$40\%) of Si is in the gas, that most (if not all) of the carbon is to be found in the gas phase.\cite{2002ApJ...579..304W,2006A&A...456..189P,2009SSRv..143..311S,2012A&A...548L...2C} This is entirely consistent with the complete destruction of a-C(:H) mantles and all carbon grains and the liberation of any a-C(:H)-incorporated Si into the gas, possibly along with some minimal silicate grain destruction. Thus, indicating that carbon is indeed a more labile element than previously thought.\cite{2008A&A...492..127S,2011A&A...530A..44J,2014A&A...570A..32B}

%------------------------------------------------------------------
\section{Dust: evolved grain mantles}
\label{sect_DustII}
%------------------------------------------------------------------ 

It now appears that core/mantle (CM) interstellar grain models seem to have re-gained ground of late\cite{2013A&A...558A..62J,2014A&A...565L...9K} after being proposed long ago as a viable model for interstellar dust.\cite{1986Ap&SS.128...17G,1997A&A...323..566L}
All of these models are based on the supposition that interstellar grain materials are mixed and that the mantles on heterogeneous core/mantle grains are perhaps the most important consequence of dust material mixing in the ISM. 
As proposed and extensively discussed in this work, it would seem that in denser regions of the ISM the grain surfaces are likely chemically-active, wide band gap, a-C:H materials with incorporated hetero-functional groups. 
This might indeed imply that, as argued here, at the outset the accreted interstellar "ice" mantles are not actually dominated by water ice but are a much more complex mix of surface-bonded organics including, for example alcohol, ketone, aldehyde and carboxylic acid surface functional groups, {\it i.e.}, $-$OH, $>$C=O, $>$C$\leqslant^{\rm H}_{\rm O}$ and $>$C$\leqslant^{\rm OH}_{\rm O}$, which may then provide a natural connection with the organic nano-globule observed in meteorites, interplanetary dust particles and cometary dust samples. The following sub-sections explore these links in detail.

%------------------------------------------------------------------
\subsection{"Organic" materials and "nano-globules"}
\label{sect_DustII_organics}
%------------------------------------------------------------------ 

The insoluble organic matter (IOM) in primitive meteorites is an interesting amalgam that exhibits significant, systematic and  interesting compositional variations that can perhaps reveal some fundamental and key information about the ISM and the origins of our solar system and of ourselves. For example, the least-heated IOM from CO carbonaceous chondrites contains fewer aromatic C=C functional groups, more nitrogen and higher ketone ($>$C=O) and carbonyl ($-$C$\leqslant^{\rm OH}_{\rm O}$) functionality,\cite{2015LPICo1856.5128D} indicating an anti-correlation between the aromatic content and oxygen-containing functional groups. This is consistent with parent body heating rendering carbonaceous matter more aromatic and, in the process, driving out molecular functional groups to eventually form a poorly graphitised type of carbon.Interestingly, there does appear to be an N-rich component in cometary organic matter that is not present in meteorites.\cite{2009LPI....40.2260D} Separate from but associated with the IOM in primitive meteorites are the so-called "organic nano-globules". 

% ORGANIC NANO-GLOBULES
\noindent \underline{\bf Organic nano-globules:} These highly intriguing spherical, core/mantle structures are ubiquitous in primitive solar system solids, exhibit fundamental compositional variations and are more aromatic than, and compositionally different from, the surrounding IOM. Based on a selection of the available literature
\cite{2006Sci...314.1439N,2008LPI....39.2391M,2009LPI....40.1130D,2009AGUFM.P14A..02D,2009_CLS_Report_DeGregorio,2015LPI....46.1609D,2015LPICo1856.5128D} the main features of the analysed organic nano-globules are summarised in the rest of this section. Possible connections between organic nano-globules and  interstellar dust are then explored within the framework of recent ideas on the nature and evolution of dust in interstellar media.

% SOURCES & DISCOVERY
\noindent \underline{\bf Parent bodies:} 
To date abundant organic nano-globules have been found in carbonaceous chondrites\cite{2006Sci...314.1439N,2006LPI....37.1455G,2008LPI....39.2391M}, chondritic porous interplanetary dust particles (IDPs)\cite{2008LPI....39.2391M} and cometary dust.\cite{2009LPI....40.1130D} 
Organic nano-globules are therefore common to both asteroidal and cometary parent bodies. 

% ISOTOPIC COMPOSITION
\noindent \underline{\bf Isotopic composition:} 
Organic nano-globules are the carriers of the most isotopically anomalous hydrogen and nitrogen components to be found in primitive materials.\cite{2006Sci...314.1439N} Almost all organic nano-globules are significantly enriched in $^{15}$N and deuterium, with respect to that of the bulk material composition, however, the N and H isotopic ratios do vary independently from globule to globule.\cite{2006Sci...314.1439N,2008LPI....39.2391M,2009LPI....40.1130D}
This latter characteristic almost certainly rules out parent body processing as an origin for the N and D isotopic anomalies, nevertheless, attached globules do have similar compositions indicating that aggregation occurred before  incorporation into the parent material.\cite{2006Sci...314.1439N} The measured isotopic anomalies within the globules are consistent with chemical fractionation in a cold medium but not with a nucleosynthetic origin because of the lack of large isotopic anomalies in carbon that would be typical of dust formed around evolved stars.\cite{2006Sci...314.1439N} 
A unique carbonaceous chondrite (Miller Range 07687), with apparently no nano-globules, does show regions with anomalous isotopic carbon compositions but these are not always associated with anomalously isotopic N and H.\cite{2015LPI....46.1609D} The majority of these isotopic anomalies are found in round, sub-micron regions but also in larger, vein-like structures (which show isotopically normal carbon).  It has been shown that radiation damage can enrich D but not $^{15}$N in electron-irradiated organics and that this also leads to the aromatisation of aliphatic compounds.\cite{2009AGUFM.P14A..02D} Interestingly, the largest $^{15}$N enrichments appear to be associated with the more aromatic nano-globules, which are more abundant in the most primitive meteorites; whereas IOM-like globules show lower, but still enhanced,  $^{15}$N enrichment.\cite{2009AGUFM.P14A..02D} 

% CHEMICAL COMPOSITION
\noindent \underline{\bf Chemical composition:} 
Organic nano-globules show rather wide variations in morphology and chemistry, in addition to variations in isotopic anomalies, which indicate multiple formation sites and different evolutionary histories.\cite{2009AGUFM.P14A..02D}
The majority of meteoritic nano-globules have similar chemistry to the IOM\cite{2009AGUFM.P14A..02D} but are, nevertheless, chemically-distinct from matrix material\cite{2006Sci...314.1439N} and tend to be more aromatic than the surrounding material.\cite{2015LPICo1856.5128D} They are generally nitrogen-rich (N/C $\sim 0.1$),\cite{2006Sci...314.1439N,2009LPI....40.1130D,2009LPI....40.2260D}  show evidence of aromatic C=C in poly-aromatic domains, nitriles ($-$C$\equiv$N)\cite{2009LPI....40.1130D}, enols ($>$C=C<$^{\rm OH}$), phenols (aromatic OH), carbonyl groups ($>$C=O in ketones, vinyl ketone $^{\rm =C<}$\hspace{-0.18cm}$>$C=O and carboxyl $-$C$\leqslant^{\rm OH}_{\rm O}$) and tend to be richer in these functional groups than the surrounding IOM.\cite{2009LPI....40.1130D,2009LPI....40.2260D,2009AGUFM.P14A..02D} However, the aromatic carbon-dominated globules show fewer carbon-oxygen function groups.\cite{2009LPI....40.1130D,2009LPI....40.2260D,2009AGUFM.P14A..02D} In particular, there is a subset of highly-aromatic nano-globules, which show no evidence for carbonyl groups but that are  $^{15}$N-anomalous, indicating an origin in the cold ISM where the $^{15}$N could have been incorporated into and preserved in aromatic domain heterocycles\cite{2009_CLS_Report_DeGregorio}, {\it i.e.}, in aromatic-rich moieties.\cite{ANT_RSOS_topdown}This is further supported by the observation that their $^{15}$N enrichment appears to be strongly associated with an insoluble macro-molecular material independent of the D-rich material.\cite{2006Sci...314.1439N} Overall, there appear to be two distinct groupings of organic nano-globules, aromatic-rich and aliphatic-rich.\cite{2002IJAsB...1..179N,2004E&PSL.224..431G,2009LPI....40.1130D} It has been noted that organic nano-globules resemble cometary CHON particles in both chemical composition and size ($20-1000$\,nm).\cite{2006Sci...314.1439N} Further, they are only found in the carbonate-free regions of meteorites suggesting that they are susceptible to oxidation.\cite{2006Sci...314.1439N} 

% STRUCTURE
\noindent \underline{\bf Structure:} 
Organic nano-globules exhibit hollow shell or filled spherical core/mantle structures, of similar composition, with diameters of $100-1000$\,nm and "mantle" thicknesses that are somewhere between $\sim 20$ and $\sim 80$\% of the particle radius.\cite{2006Sci...314.1439N,2009LPI....40.1130D}  Nano-globules are principally composed of amorphous carbon with no long-range order.\cite{2006Sci...314.1439N,2008LPI....39.2391M} In the Tagish Lake and Bells CM2 carbonaceous chondritic meteorites almost all of the nano-globules exhibit hollow spherical shell structures $\sim 70-850$\,nm in radius and shell thicknesses $\sim 100-200$\,nm, with aggregates of globules being common.\cite{2006Sci...314.1439N,2008LPI....39.2391M} An analysis of the porous, fine-grained, anhydrous cluster IDP L2005AL5\cite{2008LPI....39.2391M} revealed an interesting mineralogical assemblage, including: enstatite, forsterite, Fe-Ni sulphides, glass with embedded metal and sulphide grains (GEMS) and abundant carbonaceous material present as grain mantles, veins and spherical globules. This work further showed that the isotopically anomalous $^{15}$N hot spots were found to be associated with organic globules similar to those found in meteorites. 
Cometary organic nano-globules generally seem to be larger than meteoritic nano-globules and to have thicker walls.\cite{2009LPI....40.1130D}

% ORIGINS 
\noindent \underline{\bf Origins:} 
The $^{15}$N-rich and deuterium-rich nature of the organic nano-globules are indicative of mass fractionation under cold cloud conditions ($\simeq 10$\,K),\cite{2006Sci...314.1439N} such as in interstellar  molecular clouds or equivalently the outer regions of the solar nebula.\cite{2006Sci...314.1439N,2009_CLS_Report_DeGregorio} It has therefore been proposed that the globules were formed by the photo-processing of interstellar ices into organic refractory materials\cite{2006Sci...314.1439N} and that the hollow-shell nano-globules likely formed around now-lost, more volatile core materials or could be the result of aqueous alteration in asteroids and comets.\cite{2006Sci...314.1439N,2009_CLS_Report_DeGregorio}

% EVOLUTION
\noindent \underline{\bf Evolution:} 
It is likely that these globules experienced a wide range of thermal and chemical processing since their formation and incorporation into solar system bodies, perhaps even including some aqueous alteration.\cite{2006Sci...314.1439N}
For example, it does appear possible to form such organic nano-globule structures through the photolysis of organic-containing ices and their subsequent exposure to liquid water.\cite{2006Sci...314.1439N} However, for this to be a viable route to organic nano-globules it would require a significant presence of liquid water within comets and asteroids at some stage in their respective evolution.\cite{2009LPI....40.1130D} It has also been suggested that the organic nano-globules could have been formed by the accretion of organic matter onto icy grains and that these icy cores were latter evaporated to leave hollow shells.\cite{2009LPI....40.1130D} However, this scenario requires that icy grains form before the accretion of organic material, which seems somewhat backwards given that the organics are more refractory than and probably accrete before ices.\cite{Faraday_Disc_paper_2014,2015A&A...000A.000J,2015A&A...000A.000Y}

% ORGANIC NANO-GLOBULES AND THE ISM
\noindent \underline{\bf An interstellar dust connection?} A closer look at an albeit rather limited sampling of the organic nano-globule images available in the literature\cite{2006Sci...314.1439N,2008LPI....39.2391M,2009_CLS_Report_DeGregorio,2009LPI....40.1130D,2015LPICo1856.5128D} indicates globule central core or hole radii of $100-150$\,nm and shell or mantle depths of the order of $50-100$\,nm. It also appears that thicker mantles occur around larger cores or holes. The two analysed Stardust organic globules\cite{2009LPI....40.1130D} are significantly larger and seemingly better preserved than the thirteen or so meteoritic/IDP nano-globules in this sampling.\cite{2006Sci...314.1439N,2008LPI....39.2391M,2009_CLS_Report_DeGregorio,2015LPICo1856.5128D} This could be the result of a two-fold sample selection effect, firstly, larger and well-preserved globules are easier to detect and, secondly, larger and more robust globules would have better survived the impact with the Stardust aerogel. 

Other than the two Stardust globules the core/shell structure (core radius $\simeq 100-150$\,nm with shell thickness $\simeq 50-100$\,nm) of the organic nano-globules is remarkably similar to the core mantle structure of the large carbonaceous, a-C:H/a-C, grains (core radius $\simeq 50-300$ with mantle thickness $20-30$\,nm) in the THEMIS diffuse ISM dust model.\cite{2013A&A...558A..62J,2014A&A...565L...9K,2015A&A...577A.110Y} The evolution of these grains in denser regions of the ISM, on the outskirts of molecular clouds, is assumed to proceed via the accretion of carbon from the gas as a-C:H and therefore to the formation of thicker carbonaceous mantles in dense regions before any ice mantles accrete.\cite{2015A&A...579A..15K,2015A&A...000A.000J,2015A&A...000A.000Y} It is therefore not surprising that the pre-solar globules found in the solar system, which are samples of dense cloud matter, should have thicker mantles than the equivalent grains assumed in the THEMIS diffuse ISM model. The cycling of dust between dense and diffuse interstellar media during cycles of cloud collapse (a-C:H mantle accretion) and star formation (a-C:H mantle photo-processing to a-C) would likely lead to the accumulation of thicker a-C(:H) mantles, which would nevertheless be progressively eroded during their sojourn in the low density ISM. Thus, in the same way that silicate dust in the ISM has been processed to an amorphous form,\cite{2001A&A...368L..38D,2004A&A...420..233D} with some small fraction remaining unprocessed and crystalline,\cite{2016arXiv160102329W} carbonaceous dust would be expected to show these same traces, despite perhaps being more fragile.\cite{2008A&A...492..127S,2011A&A...530A..44J,2014A&A...570A..32B} Thus, carbonaceous core/mantle grains ought to present a range of compositions depending on their exposure to irradiation (by photons, electrons and ions) in the ISM, {\it i.e.}, aromatic-rich highly-processed and aromatic-poor (aliphatic-richer) less processed grains. Such compositions appear to be reflected in and coherent with the observed nano-globule structures. Further, this scenario is consistent with the presence of $^{15}$N enrichment which constrains the formation of nano-globules structures to low temperature regions in dense cloud environments. This nitrogen is likely to incorporated as hetero-atoms into stable and resistant poly-aromatic moieties formed in the cold ISM through accretion and low-level photo-processing.\cite{2013A&A...555A..39J,ANT_RSOS_nanoparticles,ANT_RSOS_topdown}

Another interesting aspect to emerge from the analysed organic nano-globules is that in several cases the globule outer shells or mantles show sub-grain structures with $a\lesssim 25$\,nm.\cite{2006Sci...314.1439N}\cite{2008LPI....39.2391M} Such structures look remarkably similar to the accreted and coagulated mantles in the Jones et al. diffuse ISM dust model\cite{2013A&A...558A..62J} (see their Fig.~1). Thus, the observed nano-globule mantle sub-structures, are consistent with the implied compositional and structural evolution of diffuse ISM dust in the transition towards the denser regions of the ISM. In particular, in the outer reaches of molecular clouds, a-C(:H) mantles accrete and small a-C grains ($a \lesssim 30$\,nm) coagulate onto the surfaces of the larger silicate and carbonaceous grains ($a \simeq 150$\,nm)\cite{Faraday_Disc_paper_2014,2015A&A...579A..15K,2015A&A...000A.000J,2015A&A...000A.000Y} yielding lumpy mantle structures.

%------------------------------------------------------------------
\subsection{Volatile mantles}
\label{sect_DustII_volatiles}
%------------------------------------------------------------------ 

In the cold and dense regions of the ISM, where UV photons are scarce, all grains become frosted with water-rich icy mantles as molecules progressively condense onto or form on their surfaces. 

While the laboratory assignment of ice band features to particular molecular species is secure, their use in the interpretation of constituent interstellar ice band identifications is perhaps not so completely "cut and dried".\cite{2015ARA&A..53..541B} This could in part be due to the fact that the current identifications do not, for reasons of experimental difficulty, include any materials with "polymeric" or macroscopically-bonded structure-bridging ($>$C$<^{\rm OH}_{\rm H}$, $-$O$-$, $>$C=O, \ldots) and/or structure-terminating ({\it e.g.}, $-$C$\leqslant^{\rm H}_{\rm O}$, $-$C$\leqslant^{\rm OH}_{\rm O}$, $-$C$\leqslant^{\rm O-}_{\rm O}$, O=C$<^{\rm O-}_{\rm O-}$, \ldots) functional groups. For, given the complexity of interstellar chemistry, it is envisagable that such species could also contribute to the observed IR "ice" absorption bands at some, as yet undetermined level, in at least onset ice mantle formation environments ($A_{\rm V} \lesssim 1$\,mag., $n_{\rm H} < 10^3$\,cm$^{-3}$) where grain-surface catalysed reactions must play a key role.

The classical view of volatile, molecular ({\it e.g.}, H$_2$O, CO, CO$_2$, CH$_3$OH, \ldots) ice mantle accretion by passive physisorption onto graphitic carbon and amorphous silicate grain surfaces is then most probably irrelevant to ISM studies. Instead, and in the light of the evidence and discussion presented here, at the onset of volatile mantle formation gas phase atoms and radicals almost certainly chemisorb onto and interact with activated grain surfaces. 
In the earliest stages of "ice" mantle accretion the gas interacts with nascent nano-particle surfaces, which dominate the total available grain surface in the lower-density regions of the ISM. These surfaces have been activated by reaction with atomic oxygen (nitrogen) [sulphur] to form reactive epoxide (aziridine) [episulphide] groups that form a contiguous part of the grains, {\it i.e.}, epoxylated (aziridinated) [episulphidised] grain surfaces. In somewhat denser regions, the activated surfaces will readily react with other gas phase species (principally H, O and N atoms), aided by the ambient and mildly extinguished interstellar UV radiation field (a trigger for epoxide reactions), to form surface OH (with some fraction released into the gas), ketones ($-$O$-$) and carbonyl groups ($>$C=O) in aldehyde ($-$C$\leqslant^{\rm H}_{\rm O}$), carboxylic acid ($-$C$\leqslant^{\rm OH}_{\rm O}$) and carboxylate ($-$C$\leqslant^{\rm O-}_{\rm O}$) functional groups, and perhaps also some sulphur analogues of these groups. Additionally, the reactions of aziridine functional groups could result in the formation of surface imines (=NH), amines ($-$NH$_2$) and amides ($-$C$\leqslant_{\rm O}^{\rm NH_2}$) but probably to a significantly lesser extent, or not at all, given the apparent non-observation of the NH and C$\equiv$N absorption features where CO is abundant in ice.\cite{1996ApJ...458..363W} Thus, the onset of "water ice" mantle formation, as revealed by the first appearance of a $\simeq 3.0\,\mu$m O$-$H stretching band, could be due to surface O$-$H bonds on activated carbonaceous grain surfaces, rather than in molecular water ice. However, with subsequent molecular ice mantle accretion the observed band must indeed be due to O$-$H stretching absorptions in water ice. 
Nevertheless, this $3.0\,\mu$m band profile is not fully consistent with a "pure" water ice origin because of a (well-correlated) red wing that is always present in the $3.1-3.8\,\mu$m region, and a feature/shoulder sometimes present at $\simeq 2.9\,\mu$m. The red wing can be explained by scattering from large ice grains (radii $\sim 0.5\,\mu$m) but this explanation is not entirely convincing.\cite{2015ARA&A..53..541B} However, if as proposed above, $>$C=O containing groups are present on grain surfaces prior to or at the onset of ice mantle formation, then the water ice band at $3.1\,\mu$m would naturally exhibit a broad, long wavelength wing centred at $\simeq 3.2\,\mu$m due to the IR transitions of the various carbonyl functional groups ({\it e.g.}, ketone, aldehyde, carboxylic acid, carboxylate and organic carbonate, see Fig.~\ref{fig_func_grp_IR_bands}). Currently, the prominent $6.0$, $6.85$ and $7.24\,\mu$m bands have not been unequivocally identified\cite{2015ARA&A..53..541B} but they would also appear to be consistent with an origin in a variety of carbonyl functional groups (see Fig.~\ref{fig_func_grp_IR_bands}). This would support the general idea of an origin in carboxylic acids such as HC$\leqslant^{\rm OH}_{\rm O}$.\cite{2015ARA&A..53..541B}
Further, and given the abundance of a-C:H grain materials in denser regions of the ISM (see Section \ref{sect_precepts}), the carbonyl band ought to be accompanied by a broad, $3.2-3.6\,\mu$m, aliphatic/olefinic CH$_n$ band centred at $\sim 3.4\,\mu$m (see Fig.~\ref{fig_3mic_ext}).  This hypothesis would appear to provide an entirely self-consistent explanation for the long-wavelength H$_2$O$_{\rm (s)}$ $3.1\,\mu$m ice band wing in line with the observation that the ice band and its red wing exhibit the same appearance-threshold.\cite{1996ApJ...458..363W,2015ARA&A..53..541B} This scenario is also entirely coherent with the observation of a carbonyl-containing component present in organic nano-globules and on cometary surfaces (see Section \ref{sect_DustII}) and perhaps also with the fact that the $3.1\,\mu$m ice band profile shows considerable variations between different environments.\cite{2015ARA&A..53..541B} Observations of the absorption and polarisation of the $3\,\mu$m water ice band towards the Becklin-Neugebauer object in Orion show an absorption and polarisation wing in this feature that is consistent with the presence of both water ice and a-C:H on the polarising grains \citep{1996ApJ...461..902H}. In the rest of this section this hypothesis is explored within the framework of the current interstellar ice mantle observations and the interpretation of these data.  

% FIGURE 12
\begin{figure}[!h]
\centering\includegraphics[width=5.5in]{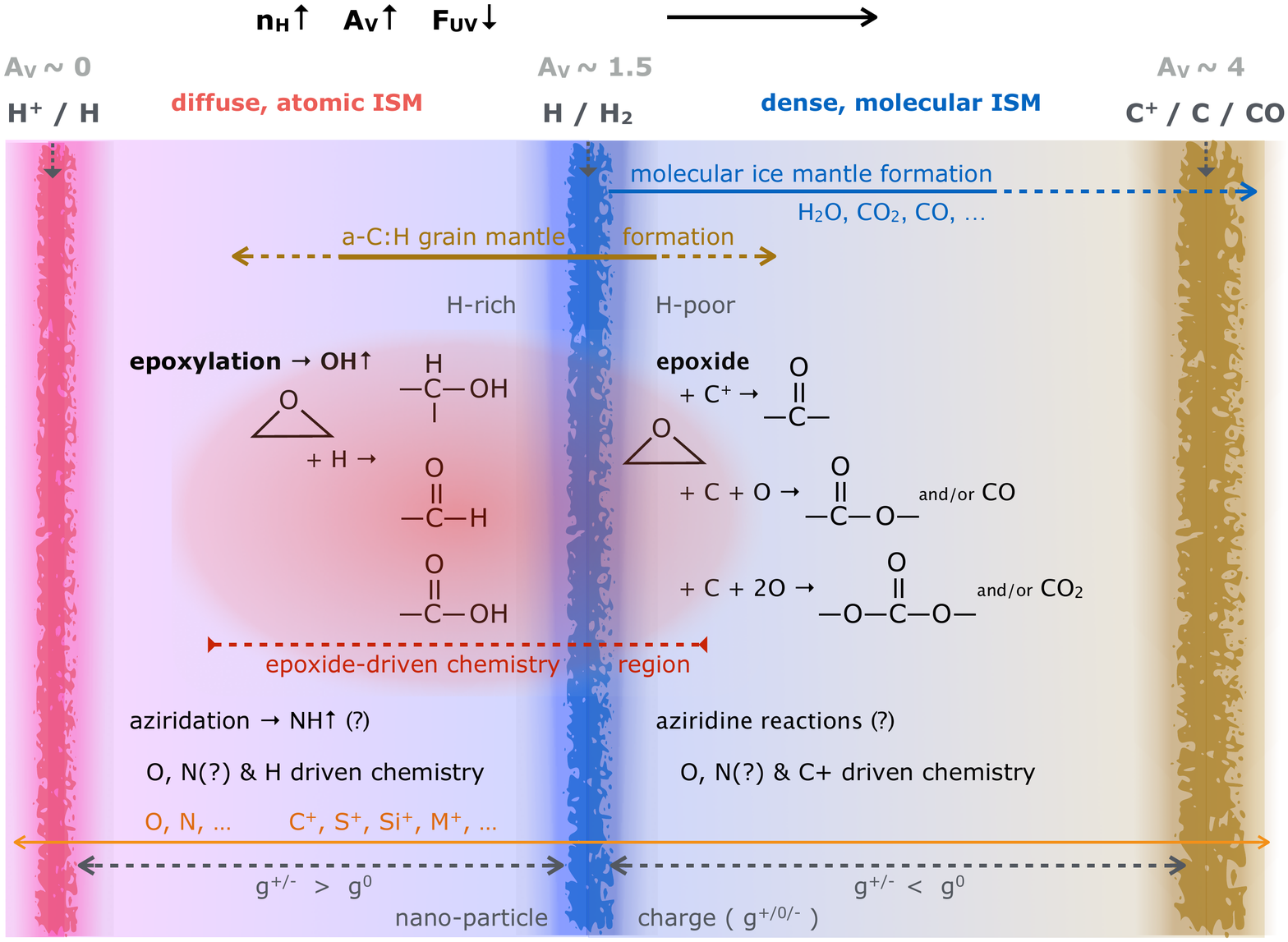}
\caption{A schematic view of the possibly important role of a-C(:H) (nano)particle surface-epoxides in driving the chemical evolutionary pathways in the transition from PDRs (PDRs) to molecular clouds.}
\label{fig_PDR_dust_chem}
\end{figure} 

In the relatively low density ISM water ice mantle formation occurs for  extinctions $A_{\rm V} \gtrsim 3.2\pm0.1$ ($\equiv n_{\rm H} \gtrsim 10^3$\,cm$^{-3}$  and $G_0 \lesssim 0.07$), equivalent to an ice formation threshold of $A_{\rm V} \simeq 1.6$\,mag. into the cloud.\cite{2015ARA&A..53..541B} but this does vary somewhat from cloud to cloud and, for instance, is lower in the Taurus molecular cloud.\cite{2001ApJ...547..872W} For CO$_2$, CO and CH$_3$OH in ice the appearance thresholds are  $A_{\rm V} \sim 1.6\pm0.1$, $\sim 3\pm1$, and $\sim 9\pm3$\,mag., respectively.\cite{2007ApJ...655..332W,2015ARA&A..53..541B} The thresholds for H$_2$O$_{\rm (s)}$ and CO$_2$$_{\rm (s)}$ in molecular-ice mantles are therefore the same,  while that for CO$_{\rm (s)}$ is somewhat higher, perhaps indicating a close chemical relationship, particularly between H$_2$O$_{\rm (s)}$ and CO$_2$$_{\rm (s)}$.\cite{2007ApJ...655..332W} The formation of H-rich methanol in ice (H/[C+O] = 2) would therefore appear to be fundamentally different from that for H$_2$O, CO$_2$ and CO in ice (H/[C+O] = 2 or 0). The observational evidence for CH$_3$OH in ice could perhaps regarded as somewhat contradictory and it is worthy of note that it and ammonia, NH$_3$, were not detected in the cometary volatiles analysed by the COSAC instrument onboard Rosetta's Phil\ae\ lander (see later), which are thought to be representative of the cold ISM ice composition. 

\begin{figure}[!h]
\centering\includegraphics[width=5.5in]{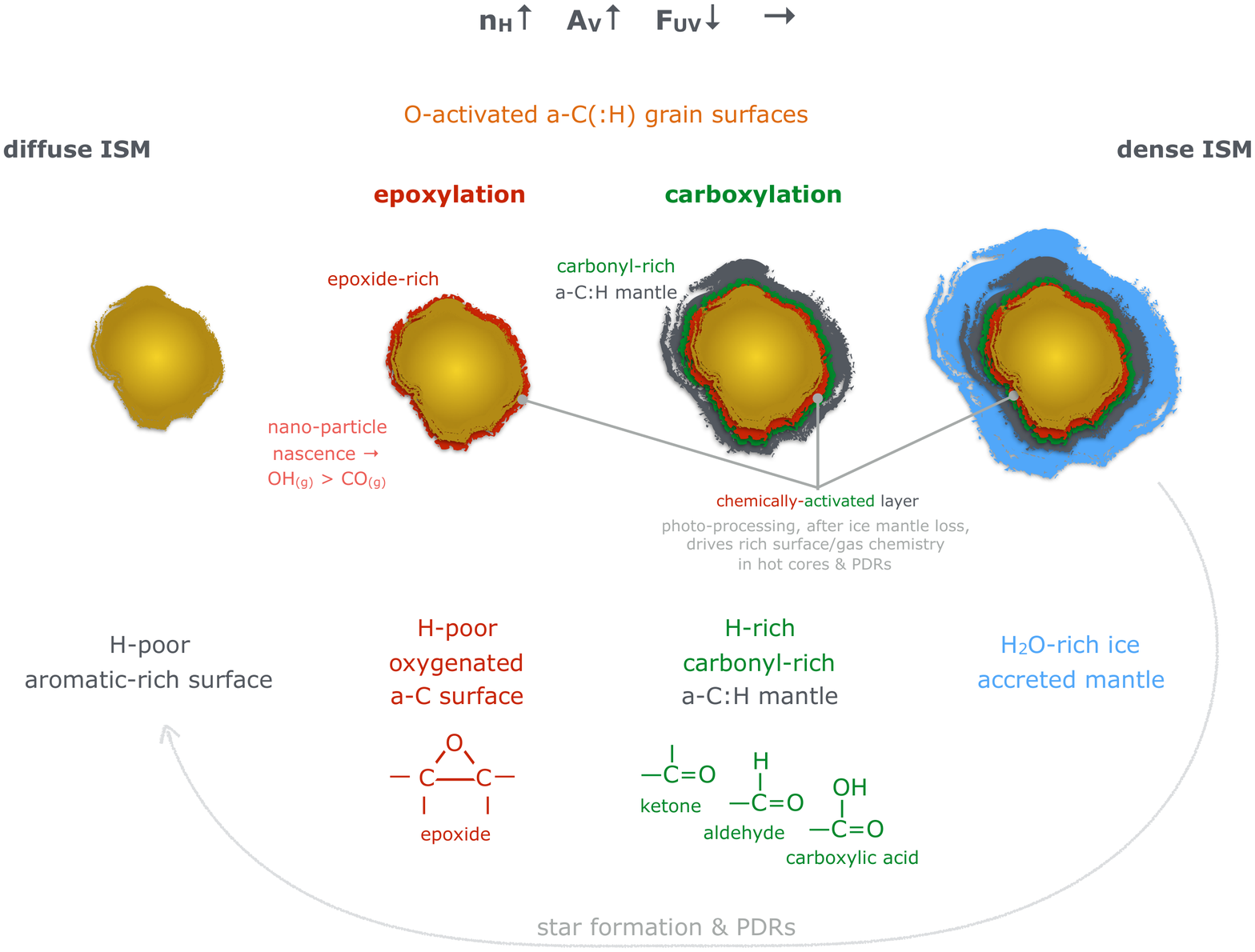}
\caption{Possible chemical evolutionary pathways for a-C(:H) dust in the transitions between the tenuous and dense regions of the ISM.}
\label{fig_dust_chem_evoln}
\end{figure}

It seems that the abundant CO$_2$$_{\rm (s)}$ observed towards Elias~16 (a quiescent dark cloud) provides something of a challenge to ice formation models because it must have formed in the absence of an embedded source and most models require UV irradiation to form CO$_2$ in an ice phase.\cite{1997ApJ...490..729W,1998ApJ...498L.159W} 
This perhaps implies a formation route via grain surface reactions and it has indeed been proposed that catalytic grain surface reactions could perhaps provide a viable route to CO$_2$ formation in the ISM.\cite{2007ApJ...655..332W,2009ApJ...695...94W} The schemas presented in Figs.~\ref{fig_epoxide_pathways}, \ref{fig_epoxide_OH_CO} and \ref{fig_dust_chem_evoln} indicate possible pathways to both water and carbon dioxide formation on epoxide-activated a-C:H grain surfaces through surface carbonyl and alcohol functional group intermediates. Such a common chemical association would provide an explanation as to why H$_2$O$_{\rm (s)}$ and CO$_2$$_{\rm (s)}$ follow one another in terms of their appearance thresholds and their preferred co-habitation in so-called "polar" ices. Further, this surface epoxide-driven mechanism would explain why CO appears to be rapidly oxidised to CO$_2$ ({\it i.e.}, CO$\downarrow$ + $-$O$-$$_{\rm (s)}$ $\rightarrow$ CO$_2$$_{\rm (s)}$: see Fig.~\ref{fig_epoxide_pathways}) in tandem with water formation at low ice column densities in the Taurus cloud [$N({\rm H_2O}) < 5 \times 10^{17}$\,cm$^{-2}$]. Additionally, it might explain why the Serpens cloud, and other enhanced gas phase CO regions, are also overabundant in CO$_2$.\cite{2007ApJ...655..332W,2009ApJ...695...94W}

Here a new four-step scenario for the onset of and early-phase formation of "ice" mantles is proposed:
\begin{enumerate}

\item The activation of a-C(:H) grain surface C=C bonds in olefinic structures (and possibly also in aromatics) by reaction with atomic O in the low-density and optically thin ISM ($N_{\rm H} \simeq 10^{20}$\,cm$^{-2}$, $A_{\rm V} \simeq 0$) to form surface epoxide groups ($>$O$_{\rm (s)}$), predominantly on carbonaceous nano-particle grain surfaces. 
In this medium the relative abundance of carbon atoms in a-C(:H) nano-particles is of the order of $N_{\rm C,np}/N_{\rm H} \simeq 1.2 \times 10^{-4}$ (120\,ppm),\cite{2013A&A...558A..62J,2015A&A...577A.110Y} The column density of carbon atoms in nano-particles is $\equiv N_{\rm C,np} \simeq 10^{16}$\,cm$^{-2}$, which implies (for an epoxide concentration of 1\% with respect to carbon) that in these regions we might expect a nano-particle surface epoxide column density $N_{\rm O,epoxide} \sim 10^{14}$\,cm$^{-2}$.

\item The reaction of epoxide-activated surfaces with gas phase hydrogen atoms under the effects of UV irradiation to form and liberate OH radicals into the gas phase. Observationally the formation of gas phase OH ($N_{\rm OH} \simeq 3 \times 10^{13}$\,cm$^{-2}$) appears to require a small degree of extinction, {\it i.e.}, $A_{\rm V} \sim 0.5$\,mag.\cite{2010MNRAS.407.2645B} 

\item In somewhat denser, more extinguished regions with $A_{\rm V} > 1$\,mag. ($N_{\rm H} \gtrsim 2 \times 10^{21}$\,cm$^{-2}$)  epoxide groups react to form abundant surface-bonded alcohol groups, which are mostly secondary, {\it i.e.}, $>$C$<^{\rm OH}_{\rm H}$, and aldehyde and carboxylic acid functional groups, {\it i.e.}, $-$C$\leqslant^{\rm H}_{\rm O}$ and $-$C$\leqslant^{\rm OH}_{\rm O}$. The alcohol groups give rise to bands at $\simeq 2.9$, 3.5, 7.3, $8.8\,\mu$m and the carbonyl functional groups to a red wing on the $3.1\,\mu$m OH band and to bands at $\simeq 3.5$, 5.8, 7.1, 8.0 and $10.6\,\mu$m (see Fig.~\ref{fig_func_grp_IR_bands}). The combination of these bands will mimic those of ice bands, which could then be mis-interpreted as arising from water ice. For example, if only 10\% of the carbon atoms in a-C(:H) nano-particles\footnote{This represents only one fifth of the surface atoms in a 1\,nm radius a-C(:H) nano-particle, which has about 50\% of its carbon atoms in the particle surface.\cite{2012A&A...542A..98J}} were to be functionalised with $-$OH and $>$C=O bonds then their nano-particle surface column density could be as high as $\simeq 2 \times 10^{16}$\,cm$^{-2}$, which is about 20\% of the typical column densities of H$_2$O$_{\rm (s)}$ CO$_2$$_{\rm (s)}$ and CO$_{\rm (s)}$ in the ices in the Taurus and Serpens molecular clouds.\cite{2009ApJ...695...94W} 
Thus, the onset formation of $-$OH and $>$C=O groups on carbonaceous nano-grain surfaces, at less than or close to the monolayer level, could seemingly make a measurable contribution to the observed "ice" bands and may even be the origin of the "polar ice" component. This should then rather be regarded as an heterogeneous growth layer on a carbonaceous (a-C:H) substrate, rather than a passively-accreted molecular ice layer. 
However, the above mentioned surface epoxide-driven processes will also lead to molecular water and carbon dioxide formation via: \\ 
$>$O$_{\rm (s)}$ + H $\rightarrow$ $-$OH$_{\rm (s)}$ followed by $-$OH$_{\rm (s)}$ + H $\rightarrow$ H$_2$O$_{\rm (s)}$ and  
$>$O$_{\rm (s)}$ + CO $\rightarrow$ CO$_{\rm 2(s)}$ \\ 
The photolysis of carboxylic acids within a solid state ice could also yield CO$_2$$_{\rm (s)}$ via: \\ 
$-$C$\leqslant^{\rm OH}_{\rm O}$$_{\rm (s)}$ + $h\nu \rightarrow$ =CH$-$$_{\rm (s)}$ + CO$_2$ 

\item Apolar H$_2$O$_{\rm (s)}$/CO$_{\rm (s)}$ ice formation would appear to be a viable follow-on process after the nano-particle surfaces have been saturated with $-$OH and C=O at the monolayer level by epoxide-driven chemistry.\footnote{It should be noted here that, in the absence of enshrouding mantles, the nano-particle surfaces will remain active (nascent) independent of whether they are "free flyers" or in aggregates.}  
At this point the reaction of CO molecules with surface epoxide groups would no longer be possible and so surface CO$_2$ formation would naturally switch off and surface-catalysed H$_2$O$_{\rm (s)}$ formation slow down.

\end{enumerate}
A schematic view of this scenario is shown in Fig.~\ref{fig_dust_chem_evoln}. 
In order to fully evaluate and quantify the above scenario it will be necessary to model this with a full chemical network approach that includes the specific surface reactivity (nascent behaviour) of a-C(:H) nano-particles. However, given that all of the relevant reaction rates, pathways and branching ratios are currently unknown, such a detailed exploration would seem to be rather premature. \\

% FIGURE 13

\noindent {\bf The CO$_{\rm 2(s)}$ formation problem in focus:} As has been highlighted an origin for formation of this ice mantle molecular component is not yet understood and is still controversial\cite{2008ApJ...678.1005P} but its formation apparently does not depend on the presence of a strong UV radiation field.\cite{1998ApJ...498L.159W,2007ApJ...655..332W,2008ApJ...678.1005P,2015ARA&A..53..541B} As this work underlines, in low-density clouds H$_2$O$_{\rm (s)}$ and CO$_{\rm 2(s)}$ appear in so-called polar ices and at higher densities this mix is associated with CO$_{\rm (s)}$ in so-called apolar ices.\cite{2008ApJ...678.1005P,2015ARA&A..53..541B} Further, it is clear that both H$_2$O and CO$_2$ are completely "frozen out" while CO still exists both in the solid and gas phases. Any viable model for the formation of CO$_{\rm 2(s)}$ in interstellar ices must be consistent with the following observational constraints:\cite{2008ApJ...678.1005P}
\begin{enumerate}

\item CO$_{\rm 2(s)}$ is present at a level of $15-40$\% with respect to H$_2$O$_{\rm (s)}$, 

\item an apparent universality of the CO$_{\rm 2(s)}$ formation route, 

\item separate CO$_{\rm 2(s)}$ molecular environments; $\simeq 85$\% with H$_2$O$_{\rm (s)}$ and $\simeq 15$\% with CO$_{\rm (s)}$ and

\item the similar $^{12}$C/$^{13}$C isotopic compositions of gas and solid phase CO and of CO$_{\rm 2(s)}$. 
\end{enumerate}
In the light of the preceding discussion it would appear that the proposed a-C(:H) grain surface epoxide-activated reactions meet all of the CO$_{\rm (s)}$ formation constraints because 
\begin{enumerate}

\item interstellar ice species are principally formed and retained on grain surfaces\cite{1984MNRAS.209..955J} and it is here proposed that CO$_{\rm 2(s)}$ and H$_2$O$_{\rm (s)}$ are formed, respectively, by the reaction of accreting gas phase CO molecules and H atoms with surface epoxide groups ($>$O) on a-C(:H) grains via: \\ 
CO$_{\rm (g)}\downarrow$ + $>$O \ $\rightarrow$ \ \ \ CO$_{\rm 2(s)}$ \ \ and \\
H$_{\rm (g)}\downarrow$ \ \ \ + $>$O \ $\rightarrow$ \ $-$OH$_{\rm (s)}$ \ \ followed by \ \ H$_{\rm (g)}\downarrow$ + \ $-$OH$_{\rm (s)}$ \ $\rightarrow$ \ H$_2$O$_{\rm (s)}$. \\
In a gas with $n_{\rm H} \geqslant 100$\,cm$^{-3}$ CO is present in the gas at a level $\simeq 20$\% with respect to O$_{\rm (g)}$.\cite{2010ApJ...710.1009W} Thus, if the above epoxide-driven formation reactions leading to CO$_{\rm 2(s)}$ and H$_2$O$_{\rm (s)}$ proceed with equal efficiency, upon accommodation onto grain surfaces, then CO$_{\rm 2(s)}$ would naturally be present within ices at a level of $\sim 20$\%, with respect to H$_2$O$_{\rm (s)}$.  

\item a-C(:H) grain surface epoxide-driven chemistry relies on the presence of reactive surface groups that are activated by relatively mild UV radiation conditions. This reaction pathway is then likely to be widespread and universal in both its nature and operation.  

\item With the proposed reaction pathway it seems that the initial stages of ice formation occur onto reactive surfaces and quite naturally lead to an initial "polar" ice composition in which CO$_{\rm 2(s)}$ formation is directly related to, and goes hand-in-hand with, H$_2$O$_{\rm (s)}$ formation.  When the reactive surface and near surface layers are saturated CO$_{\rm 2(s)}$ formation will switch off and the accreting CO can no longer be oxidised but simply accretes with H$_2$O$_{\rm (s)}$ and any remaining CO$_2$ from the gas. Thus, early stage "polar" ice formation is driven by an active chemistry phase and late stage "apolar" ice formation by gas phase freezing onto benign surfaces. 

\item By this mechanism CO$_{\rm 2(s)}$ forms directly from the interaction of gas phase CO with a-C(:H) grain surface epoxide groups. Hence, the $^{12}$C/$^{13}$C isotopic compositions of gas and solid phase CO and of CO$_{\rm 2(s)}$ are directly related and can only be similar. 

\end{enumerate}
All of the above are then entirely in agreement with the constraints imposed by the observations of CO$_{\rm 2(s)}$ in the ices seen in the outer reaches of molecular clouds and around low mass proto-stars. \\

\noindent {\bf "Methanol" in focus, CH$_3$OH$_{\rm (s)}$ or 2$^\circ$ alcohols ($>$C$<^{\rm OH}_{\rm H}$)?:} It is somewhat troubling in interstellar ice and gas chemistry studies that methanol appears to show orders of magnitude variation in its ice phase abundance and that it is also way more abundant than that required to explain the observed gas phase methanol abundances.\cite{2011ApJ...742...28W} Further, the solid methanol, CH$_3$OH$_{\rm (s)}$, IR features appear to be present in only a few percent of lines of sight through dark clouds and towards young stellar objects.\cite{2013ApJ...774..102W} A simple solution to this conundrum would be if methanol formation (by the proposed hydrogenation of accreted gas phase CO) is not as efficient as assumed but instead that the IR bands attributed to methanol are, in no small part, due to secondary alcohol functional groups on grain surfaces, which can exhibit sufficiently-confusing IR spectra (see Fig.~\ref{fig_func_grp_IR_bands}).

%------------------------------------------------------------------
\subsection{Photolysis effects}
\label{sect_DustII_photolysis}
%------------------------------------------------------------------ 

As recently proposed the outer carbonaceous layers of grains, be they a-C(:H) grains or mantles on amorphous silicates, will likely be converted to a-C materials via UV/EUV photolysis.\cite{2012A&A...540A...1J,2012A&A...540A...2J,2012A&A...542A..98J,2013A&A...558A..62J} The photolysis of hetero-atom doped a-C(:H) materials, {\it i.e.}, a-C:H:O:Ns, therefore leads to the loss of the more labile aliphatic components and to the aromatisation of the outer layers but perhaps to the retention of the more resistant carbonyl-containing functional groups, such as $>$C=O, $>$C$\leqslant^{\rm H}_{\rm O}$ and $>$C$\leqslant^{\rm OH}_{\rm O}$, which may be at the origin of the broad $\sim 3.2\,\mu$m carboxyl band in comet 67P reflectance spectrum and also the ubiquitous red wing on the interstellar $\sim 3\,\mu$m water ice band. Such a scenario might also provide an explanation for the aromatic-rich organic nano-globules with a significant carbonyl component. Hard UV/EUV photolysis of such globules could remove even this carbonyl component and so explain the carbonyl-poor organic nano-globules, if oxygen-containing functional groups are preferentially removed but nitrogen hetero-atoms are retained in more resistant five-fold aromatic ring systems. It also appears that interstellar radiation field-driven chemistry in PDRs could lead to the formation of small organic molecules and radicals in these regions, {\it e.g.}, C$_2$H, C$_3$H, l-C$_3$H$^+$, c-C$_3$H$_2$ , l-C$_3$H$_2$, C$_4$H and CH$_3$CN,\citep{2005A&A...435..885P,2012A&A...548A..68P,2015ApJ...800L..33G} C$_n$H$_m$X$_p$ species and even molecular hydrogen.\cite{2012A&A...540A...2J,2012A&A...542A..98J,2013A&A...558A..62J,2014A&A...569A.119A,2015A&A...581A..92J,2015A&A...584A.123A}

%------------------------------------------------------------------
\subsection{Hot core processing}
\label{sect_DustII_hotcore}
%------------------------------------------------------------------ 

% TABLE 
\begin{table}[!h]
\caption{A partial inventory of HCO species detected in hot cores and their relative abundances\cite{1998A&A...337..275N} in order of magnitude groupings.}
\label{table_hot_cores}
\begin{center}
\begin{tabular}{llccccc}
\hline
                                                 &                   &                 &                           &                                 &                          \\[-0.25cm]
          &               & Relative     & No. of        &   No. of     &  No. of      &                                                           \\
          &               & abundance & H atoms    &   C atoms  &  O atoms  &                                                          \\
Name & Formula & $\times 10^{10}$  &( $N_{\rm H}$ ) & ( $N_{\rm C}$ ) & ( $N_{\rm O}$ ) & $\frac{N_{\rm H}}{(N_{\rm C}+N_{\rm O})}$) \\ 
         &               &                          &                           &                          &                                                            \\[-0.3cm]
\hline
          &       &      &      &      &       &        \\[-0.25cm]  
methanol          &  CH$_3$OH                                 & $800-2000$ &   4   &  1    &  1     &   2     \\[0.05cm]
          \hline
        &       &      &       &        &       &       \\[-0.25cm]
dimethyl ether  & CH$_3$OCH$_3$                        & $100-300$  &   6   &   2   &   1    &    2    \\[0.05cm]
ethanol             & CH$_3$CH$_2$OH                     & $40-200$    &   6   &    2  &    1   &    2    \\[0.05cm]
          \hline
        &       &      &       &        &       &       \\[-0.25cm]
ethanal             & CH$_3$C$\leqslant^{\rm H}_{\rm O}$ & $10-30$       &   4   &   2   &   1    &    $\frac{4}{3}$    \\[0.05cm]
methanoic acid & HC$\leqslant^{\rm OH}_{\rm O}$        &  $9-10$        &    2  &   1   &    2   &     $\frac{2}{3}$   \\[0.05cm]
ethylene oxide  & $^{\rm H}_{\rm H}>$C$_-^{\rm O}$C$<^{\rm H}_{\rm H}$ & $2-6$ &   4   &   2   &   1    &    $\frac{4}{3}$    \\[0.05cm]
          \hline
        &       &      &       &        &       &       \\[-0.25cm]
\end{tabular}     
\end{center} 
\end{table}

To date it seems that the molecule ethylene oxide ($^{\rm H}_{\rm H}>$C$_-^{\rm O}$C$<^{\rm H}_{\rm H}$) has only been detected towards a small number of hot core sources.\cite{1997ApJ...489..753D,1998A&A...337..275N} 
Its distribution therefore appears to be limited to rather particular environments, {\it i.e.}, compact ($< 0.5$\,pc) hot cores within massive star forming regions that exhibit a high abundance of large saturated molecules.\cite{1997ApJ...489..753D,1998A&A...337..275N} Here volatile ice mantles have presumably been evaporated into the gas phase in the regions immediately surrounding the young massive stars. It appears that current interstellar and circumstellar chemistry models are not yet able to explain the observed gas phase abundances of species such as ethanol, methanol and ethylene oxide in these objects.\cite{1997ApJ...489..753D} A partial inventory of gas phase molecules in a number of hot core sources\cite{1998A&A...337..275N}, as shown in Table~\ref{table_hot_cores} in decreasing abundance sequence, indicates that the three least abundant species contain an epoxide ring or a carbonyl bond ($^{\rm H}_{\rm H}>$C$_-^{\rm O}$C$<^{\rm H}_{\rm H}$, HC$\leqslant^{\rm OH}_{\rm O}$ and CH$_3$C$\leqslant^{\rm H}_{\rm O}$) and a hydrogen to heavy atom ratio close to unity, $N_{\rm H}$/($N_{\rm C}$+$N_{\rm O}$) $= 1.0\pm0.3$, whereas the three significantly more abundant molecules (CH$_3$CH$_2$-OH, CH$_3$-O-CH$_3$ and CH$_3$-OH) all have a value of $N_{\rm H}$/($N_{\rm C}$+$N_{\rm O}$) exactly equal to two. This perhaps indicates formation via the hydrogenation from more primitive (H-poorer, grain surface) epoxide- and carbonyl-containing species in a hydrogen-rich (gas phase) environment. 
Indeed it would appear possible to form most of the molecules in Table~\ref{table_hot_cores} by the reaction of "root" epoxide functional groups on and/or within grains at the epoxide activated carbonaceous grain surface layers just underlying volatile ice mantles (see the above sub-section \ref{sect_DustII_volatiles}). With this scenario, it is to be expected that the most reactive root epoxides (at the activated grain-ice mantle interface) ought then, as observed, to be the least abundant because of their rapid reaction under the combined effects of ice mantle evaporation, grain heating, gas phase molecule-grain interactions and UV stellar radiation to form progressively more hydrogenated species on grain surfaces that are then released into the gas. This appears to be borne out by the observed rotational temperatures of the molecules listed in Table~\ref{table_hot_cores}, which generally increase from bottom to top, indicating a core/envelope structure within these sources. 
The envelope ($\phi \simeq 0.3$\,pc, $T_k = 40-60$\,K and $n_{\rm H} = 10^5-10^6$\,cm$^{-3}$) appears to favour the formation of the more "primitive" ethylene oxide, ethanal and methanoic acid molecules while the core ($\phi \simeq 0.06$\,pc, $T_k = 100-200$\,K and $n_{\rm H} = 10^6-10^8$\,cm$^{-3}$) favours the more "evolved", hydrogen-richer ethanol and methanol, while dimethyl ether appears under both core and envelope conditions.\cite{1998A&A...337..275N} 
The scenario is then of a central, dense and hot core around a young star that expands into the nebular cloud from which it formed. The surrounding cloud material is then progressively ablated and heated, evaporating ice mantles and so triggering the underlying epoxylated grain surfaces to drive both the grain surface and gas phase chemistry.
Thus, epoxide-driven grain surface reactions could be the initiating driver for the complex chemistry that is observed under the particular  conditions existing within massive young star hot core environments, ultimately resulting in the abundant, hydrogen-rich species methanol, dimethyl ether and ethanol. 

In summary, it appears that the effects of an active grain (nano-particle) surface chemistry in star forming regions will need to be included into circumstellar and interstellar chemistry models.

%------------------------------------------------------------------
\subsection{Comets and chemistry}
\label{sect_comets}
%------------------------------------------------------------------

Comets are repositories of primitive interstellar and solar nebula matter and therefore carry important information on our origins. With the arrival of Rosetta/Philae at Comet 67P/Churyumov-Gerasimenko we have gained a deeper insight into these primitive solar system bodies. 
For example, the comet chemistry, as measured by the Cometary Sampling and Composition (COSAC) instrument detection of sixteen organic molecules, shows a large abundance of carbonyl ($>$C=O) species; six of the sixteen molecules contain carbonyl bonds, including: 
3 aldehydes (R-C$\leqslant^{\rm H}_{\rm O}$, with R = CH$_3$, CH$_2$OH and CH$_3$CH$_2$), 
2 amides (R-C$\leqslant^{\rm NH_2}_{\rm O}$ with R = H and CH$_3$) and 
1 ketone, (CH$_3$)$_2$C=O.\cite{2015Sci...349b0689G} 
However, the real surprise is what was not detected, {\it i.e.}, no ammonia, NH$_3$, methanol, CH$_3$OH, carbon dioxide, CO$_2$, or carboxylic acids, R-COOH, were present in measurable quantities.
Half of the sixteen detected species contain C=O double bonds (in both $>$C=O and =C=O bonds) and half contain nitrogen atoms in -NH$_2$ amines (4/16), -N= groups (2/16) and -C$\equiv$N nitrile groups (2/16). 
However, perhaps most surprising of all, the most abundant molecule, after water, is formamide, H-C$\leqslant _{\rm O}^{\rm NH_2})$, which was detected at the $\sim 2$\% level relative to water. 

Taking the abundance numbers from the published COSAC results\cite{2015Sci...349b0689G}, including water, the effective stoichiometry of the sixteen detected cometary organics ({\it c.f.}, the cosmic elemental ratios) is $\simeq$ H$_{230}$O$_{107}$C$_{14}$N$_{6}$, equivalent to C/O = 0.13 ($\simeq 0.70$), N/O = 0.06 ($\simeq 0.20$) and N/C = 0.43 ($\simeq 0.29$). 
Thus, with respect to oxygen, the detected organics are deficient by about $70-80$\% in both carbon and nitrogen, which implies that within the comet there must be, or originally must have been, a more carbon-rich phase with N/C $\simeq 0.25$ that likely also contains some oxygen. 
This sort of composition does resemble the above-discussed organic nano-globules that, as per the sixteen detected organics, also contain carbonyl (ketone), nitrile and alcohol functional groups but also contain carboxyl groups, aromatic domains and are relatively nitrogen-rich (N/C $\lesssim 0.1$). 
Further, such a carbonyl/carboxyl/hydroxyl-containing material, also likely containing aliphatic and/or aromatic CH$_n$ bonds, as well as H$_2$O, NH/NH$_2$ and NH$_4^+$ groups/molecules,\footnote{{\it N.B.}, The contribution of water ice is likely to be rather low because the $3.1\,\mu$m ice band is lacking in the spectra and, further, nitrogen may not be so abundant because the NH$_n$ absorption profiles do not fit the $\sim 3.2\,\mu$m band particularly well.\cite{2015EPSC...10..621Q}} appears to be consistent with the broad absorption band centred at $\sim 3.2\,\mu$m\cite{2015LPI....46.2092Q}, which was observed in the reflectance spectrum of Comet 67P/Churyumov-Gerasimenko made with the Visible InfraRed Thermal Imaging Spectrometer (VIRTIS) instrument on Rosetta.\cite{2015EPSC...10..621Q,2015LPI....46.2092Q,2015Sci...347a0628C}

If such a broad $\sim 3.2\,\mu$m absorption band were to be present in the interstellar dust towards denser regions it ought to appear before the onset of the water ice $\sim 3.1\,\mu$m absorption band. 
Such a band could be formed by atomic O (and N) reactions on interstellar a-C(:H) grain surfaces leading to epoxide (and aziridine) functional groups that further react with other gas phase species during mantle accretion ({\it e.g.}, the sticking of C, O, CO, OH, \ldots) to form carbonyl/carboxyl/hydroxyl-containing carbonaceous shells much akin to the analysed organic nano-globule structures. 
Thus, if present, a $\sim 3.2\,\mu$m absorption "carbonyl" band could perhaps give rise to the impression of early ice mantle accretion if it were to be mis-identified as the water ice absorption feature. 
The observed organic nano-globules (see sub-section \ref{sect_DustII_organics} above) are rather rich in carbonyl bonds and therefore ought to exhibit a broad $\sim 3.2\,\mu$m absorption feature. The strong indication that they derive from cold (and therefore dense media) would seem to lend support to the idea that the presence of such a feature in interstellar dust in the outer reaches of molecular clouds should not be excluded. 

The Comet 67P refractory grains directly-analysed with the COmetary Secondary Ion Mass Analyzer (COSIMA) dust particle instrument onboard Rosetta appear to be somewhat reminiscent of IDPs in that they contain olivine and pyroxene silicates, iron sulphides but no clear evidence for organic matter.\cite{2016ApJ...816L..32H} The latter result must be due to a limited sampling because we know from the Stardust results, from IDPs and other Rosetta instrument measurements that comets contain significant amounts of organic matter. 
For example, the VIRTIS reflectance spectra also appear to show significant sub-structure at the positions of the aromatic and aliphatic CH$_n$ absorption bands in all of the spectra, at $\simeq 3.3\,\mu$m and $\simeq 3.4\,\mu$m, respectively.\cite{2015Sci...347a0628C}
Thus, the analysis of the Comet 67P VIRTIS reflectance spectrum clearly indicates the presence of non-volatile organic macro-molecular materials.\cite{2015EPSC...10..621Q,2015LPI....46.2092Q,2015Sci...347a0628C} 
Further, the dark surface of Comet 67P, with a normal albedo of 6\%, is typical of small solar system bodies and appears to be consistent with dark refractory organics containing poly-aromatic organics mixed with sulphides and Fe/Ni alloys.\cite{2015LPI....46.2092Q}

%------------------------------------------------------------------
\section{Testing the icy waters}
\label{sect_tests}
%------------------------------------------------------------------

Clearly it would be possible to test the viability of many of the above scenarios by dedicated low-temperature ($T \simeq 20$\,K) interstellar analogue laboratory experiments and it is hoped that the ideas presented here might inspire some such efforts in this direction, such as:  
\begin{itemize}
\item the irradiation of an a-C:H thin film with atomic oxygen (nitrogen) [sulphur] at low temperatures to test the degree of surface epoxylation (aziridation) [episulphidation] that is possible. To this end an a-C:H film with $\sim$50\% atomic hydrogen would likely be a good starting material, and the most reactive, because it contains a high olefinic, C=C, bond fraction,\cite{2012A&A...540A...1J,2012A&A...540A...2J}
\item a targeted study of the reaction of H atoms and CO molecules at epoxylated a-C:H surfaces with the aim of exploring the possible formation of water-less and CO$_2$-less "polar ice" analogues containing only surface-bonded alcohol, $-$OH, and carboxyl, $-$C$\leqslant^{\rm O-}_{\rm O}$, functional groups and 
\item a general investigation of the interaction of epoxylated a-C:H surfaces with H, O, N, C and S atoms and H$_2$, CO and CO$_2$ molecules with the aim of spectroscopically-identifying {\em surface} $-$OH, $-$O$-$, $>$C=O, $-$C$\leqslant^{\rm H}_{\rm O}$, $-$C$\leqslant^{\rm OH}_{\rm O}$, $-$C$\leqslant^{\rm O-}_{\rm O}$, $-$C$\leqslant _{\rm O}^{\rm NH_2}$,   {\it etc.} functional groups,   
{\it i.e.}, alcohols, ethers, and carbonyl-containing ketones, aldehydes, carboxylic acids, carboxylates and amides based on the above predicted reaction pathways.  
\end{itemize}
It is extremely unlikely that such an experimental investment would be for nought because, at worst, interesting things always come from the unexpected and, at best, the experimental results might just indicate the viability of some of the ideas presented here.

%------------------------------------------------------------------
\section{Summary and conclusions}
\label{sect_conclusions}
%------------------------------------------------------------------

In the light of the explorations presented here it appears that the physics and especially chemistry of interstellar dust is much more complicated than has previously been given consideration. In particular, it now appears that the carbonaceous dust component is much more responsive to its environment than the silicate dust.  
For example, the rather variable carbon depletions in the ISM indicate that the carbon within a-C(:H) grains is a rather labile dust element. The re-accretion of carbon  in the denser ISM, resulting in the formation of a-C(:H) carbon mantles on all grains, would then seem to be requisite and a natural consequence of its "volatility". 
Core/mantle-type grain and aggregate structures would then appear to be the norm and such structures are reflected in the organic nano-globules extracted from meteoritic and cometary materials, which were clearly formed in a cold interstellar environment. 
Indeed, interstellar core/mantle dust models appear to be rather successful in explaining the dust observables, especially effects such as C-shine, and are therefore re-gaining acceptance after first being proposed more that thirty years ago by the late Mayo Greenberg. The latest core/mantle model, THEMIS, is built upon this early foundation but has added a more physically-realistic nano-particle physics and surface chemistry into the mix.   

One particularly new, innovative and promising idea presented here is the possible role of grain surface epoxylation ($>$C$_-^{\rm O}$C$<$) and aziridination ($>$C$_-^{\rm N}$C$<$) in driving interstellar chemistry in diffuse-dense ISM interface regions. 
This surface functionalisation should occur as a result of O and N atom interaction with olefinic-rich a-C:H grains through: \\ \\ 
O$_{\rm (g)}\downarrow$ \ \ + \ \ $>$C=C$<_{\rm (s)}$ \ \ $\rightarrow$  \ \ $>$C$_-^{\rm O}$C$<_{\rm (s)}$ \hspace{1.0cm}  epoxylation \ \ \ \ and \\ \\
N$_{\rm (g)}\downarrow$ \ \ + \ \ $>$C=C$<_{\rm (s)}$ \ \ $\rightarrow$  \ \ $>$C$_-^{\rm N}$C$<_{\rm (s)}$ \hspace{1.0cm}  aziridination. \\ \\ 
It is apparent that nano-particle surface epoxide formation and reaction (and that of the analogous aziridine and episulphide groups), to form ketone, aldehyde, carboxylic acid, carboxylate and organic (poly)carbonate functional groups on grain surfaces, could provide a coherent, connected and self-consistent solution to some current interstellar chemistry conundrums, including: 
\begin{itemize}
\item the formation of OH in the tenuous ISM, 
\item anomalous oxygen depletion, 
\item the depletion of sulphur in the denser ISM, 
\item the nature of the CO dark gas, 
\item the formation of "polar ice" mantles, 
\item an explanation for the red wing on the 3\,$\mu$m water ice band, 
\item the basis of the O-rich chemistry in the hot cores around massive young stars,  
\item the origin of organic nano-globules and
\item the $\sim 3.2\,\mu$m "carbonyl" absorption band observed in comet reflectance spectra. 
\end{itemize}
This possible role of epoxide structures on interstellar grain surfaces might therefore merit further observational, experimental and modelling explorations.

\section*{Acknowledgment}
Thank you to everyone and, especially, to those who had the fortitude to read this series of papers. 
The ideas presented in this study have been triggered by disparate discussions with so many people that, 
seemingly, everyone who I have ever discussed dust with has left at least a small grain of an idea that has blossomed into the ideas presented here. 
Nevertheless, the author is perfectly happy to assume full responsibility for all of the wide-ranging notions, speculations and any daft ideas presented here.

%%%%%%%%%% Insert bibliography here %%%%%%%%%%%%%%

%% for the bibliography, at the end
\footnotesize{
\bibliography{Ant_bibliography} % your references Yourfile.bib
\bibliographystyle{rsc} 
}

\end{document}